\documentclass[fleqn]{article}
\usepackage{graphicx}

\begin{document}

\title{
{\sf H\"older Inequalities and Isospin Splitting of the Quark Scalar Mesons} }

\author{
Fang Shi, T.G. Steele
\thanks{email: steelet@sask.usask.ca}
\\
{\sl Department of Physics and Engineering Physics}\\
{\sl University of Saskatchewan}\\
{\sl Saskatoon, Saskatchewan S7N 5E2, Canada.}\\[10pt]
V.Elias
\thanks{email: velias@julian.uwo.ca}
, K.B. Sprague, Ying Xue\\
{\sl Department of Applied Mathematics}\\
{\sl University of Western Ontario}\\
{\sl London, Ontario N6A 5B7, Canada.}\\[10pt]
A.H. Fariborz\thanks{email: amir@suhep.phy.syr.edu}\\
{\sl Department of Physics}\\
{\sl Syracuse University }\\
{\sl Syracuse, New York 13244-1130, USA}
}
\maketitle
\begin{abstract}
A H\"older inequality analysis of the QCD Laplace sum-rule which probes the  non-strange ($n\bar n$)
components of the  
$I=\{0,1\}$ (light-quark) scalar  mesons supports the methodological 
consistency  of an effective continuum contribution from 
instanton effects. This revised formulation  enhances the magnitude of the instanton contributions
which split the degeneracy between the $I=0$ and $I=1$ channels.  
Despite this enhanced isospin splitting effect,
analysis of the  Laplace and finite-energy sum-rules seems to preclude identification of 
$a_0(980)$ and a light broad $\sigma$-resonance
as states with predominant $n\bar n$ components, in which case their dominant components
could then be meson-meson ($\pi\pi$ or $K\bar K$), multiquark, or glueball states.
This apparent decoupling of $\sigma\left[\equiv f_0(400-1200)\right]$ and  $a_0(980)$ from the
quark $n\bar n$ scalar currents suggests 
the possible identification of the $f_0(980)$ and $a_0(1450)$ as the lightest $I=\{0,1\}$ 
scalar mesons in which the $n\bar n$ component is dominant.   
\end{abstract}

\section{Introduction}\label{intro_sec}
The nature  of the scalar mesons is a challenging problem in hadronic physics.  
A variety of interpretations exist for the lowest-lying scalar resonances 
($\sigma$ or $f_0(400-1200)$, $f_0(980)$, $f_0(1370)$, $f_0(1500)$, $a_0(980)$, $a_0(1450)$ \cite{PDG})
including conventional quark-antiquark ($q\bar q$) states, $K\bar K$ molecules, gluonium, four-quark models, 
and dynamically generated thresholds \cite{interp,Speth,Jaffe}. 
The nature of the $f_0(400-1200)$ is particularly important because of its
possible interpretation as the $\sigma$ meson of chiral symmetry breaking.

In previous work we have used QCD Laplace sum-rules to study the various possibilities for the lowest-lying non-strange
scalar mesons \cite{first_paper}.  An important component of this analysis is the inclusion of instanton effects,
which are known to be present in the scalar and pseudoscalar channels \cite{instanton,ins_liquid}.  Instanton contributions 
represent finite correlation-length QCD vacuum effects, and are the only known theoretical mechanism that 
distinguishes between the $I=0$ and $I=1$ channels in the presence of $SU(2)$ flavour symmetry \cite{isobreaking}.
The analysis of ref. \cite{first_paper} showed that the isoscalar state cannot be simultaneously light and wide, and that the 
mass scale of the isovector state lies significantly above $1\,{\rm GeV}$.   These results suggested 
tentative identification of the $f_0(980)$
and $a_0(1450)$ as the lightest $q \bar q$ scalar mesons, with instanton effects being solely 
responsible for this large mass splitting between isopartners.   

In this paper we employ the H\"older inequality technique \cite{holder} to test the theoretical validity of the $I=\{0,1\}$
QCD Laplace sum-rules for the non-strange scalar currents with instanton effects included. 
Such an analysis is motivated by the  large
instanton-generated $I=\{0,1\}$ splitting discussed above.  The inequality analysis confirms that the instanton effects have an effective 
continuum contribution \cite{instanton_continuum}, leading to  revised expressions for instanton contributions to the 
sum-rule.  As will be discussed below, the revised instanton formulation leads to further
enhancement of instanton effects, suggesting  increased splitting between the 
$I=\{0,1\}$ channels. 
Thus, the inclusion of an instanton contribution to the continuum  
provides the motivation for 
 a revised QCD sum-rule analysis of the $I=\{0,1\}$ 
non-strange scalar mesons.       

Our analysis will be restricted to the non-strange ($n\bar n$) currents since we believe that a full analysis including
mixing with the $s\bar s$ scalar current, as presumably occurs for the observed ($I=0$) hadronic states, is beyond
the scope of a reliable sum-rule analysis.  However, our final results can be viewed as 
predictions for the primitive (unmixed) $n\bar n$ states which can then be used as constraints or support for studies of the 
structure of the  $q\bar q$ scalar nonet \cite{nonet}.  
%Alternatively, the sum-rule analysis can be viewed as a probe of the $n\bar n$ component of  observed  hadronic states.

 The possible meson-meson content of the observed scalar resonances is also problematic
within the context of sum-rule methodology. A sum rule analysis of colourless (scalar)
combinations of four-quark operators is a formidable challenge, since the leading perturbative
terms are already a two-loop effect in the  sum rule. However, the $n\bar n$
sum-rule analysis presented here does have value as a direct probe of the explicit $n\bar n$ 
content of
observed hadronic states.  The failure to 
observe a strong $n\bar n$ sum-rule signal for a known resonance indicates that such a state has minimal $n\bar n$ content, and is hence
not a strong candidate for a $n\bar n$ member of a $q\bar q$ nonet.  An approach for assessing the $n\bar n$ component of the known hadronic states is discussed in Section \ref{multi_res_sec}.

Laplace and finite-energy sum-rules will be used in our present analysis.  Section \ref{fesr_mot_sec}  demonstrates that the
lowest two finite-energy sum-rules (FESR's) are impervious to the resonance width(s) for a wide class of resonance models, and establishes
criteria for the existence of a light resonance masked by the contributions of heavier resonances.  Section
\ref{fesr_theor_sec} provides the theoretical expressions for the FESR's and examines the constraints on a hidden light
$\sigma$.  Section \ref{laplace_theor_sec} provides expressions for the lowest Laplace sum-rule, and 
demonstrates that the instanton continuum leads to an {\em enhancement} of instanton effects.  
Consistency of the 
Laplace sum-rule with fundamental inequalities is shown in Section \ref{holder_sec} to be upheld 
after inclusion of the instanton continuum contribution, and bounds on the parameter space of 
sum-rule validity are established.  Phenomenological predictions for the $I=\{0,1\}$ scalar resonances 
are obtained from the Laplace sum-rules in Section \ref{phenom_lap_sec}.  
These predictions consider the effects of  different resonance models, alternative analysis techniques, and higher-loop 
corrections.  Conclusions are presented in Section \ref{conc_sec}.

\section{Phenomenological Motivation for Finite Energy Sum Rules}
\label{fesr_mot_sec}
We seek to employ QCD sum-rule methods to determine whether a $q\overline{q}$
interpretation is possible for the light ($m \sim 500 - 600 MeV$), 
broad ($\Gamma \sim 350 -700 MeV$) $\sigma$-resonance suggested 
particularly, but not exclusively \cite{Augustin}, by phenomenological re-analysis 
of $\pi-\pi$ scattering phase shifts \cite{Sannino,lightsigma}.  Such analysis is complicated, 
of course, by the very large width of the resonance, which renders inappropriate 
the narrow-resonance approximation characteristic of virtually all sum-rule analyses.  
One is faced with the choice of incorporating the resonance shape directly into the 
phenomenological side of the sum rules, or alternatively, of utilizing a set of sum rules whose 
phenomenological content is insensitive to resonance-width effects.  In this section and the 
section that follows, we will embrace the latter approach via utilization 
of the first two finite energy sum rules (FESR's) in the scalar current channel. 

\begin{figure}[hbt]
\centering
\includegraphics[scale=0.5]{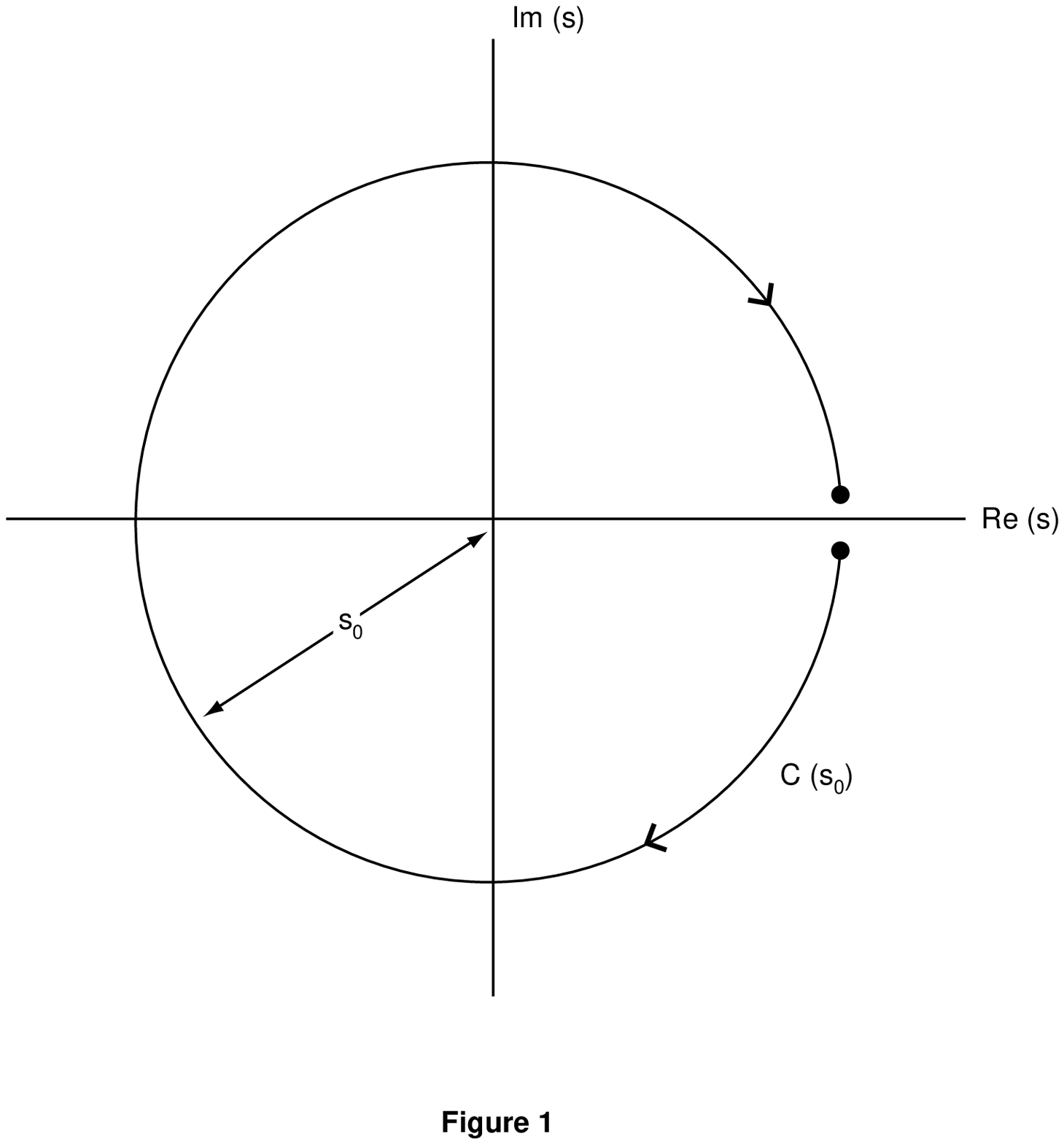}
\caption{
The contour $C\left(s_0\right)$.
}
\label{fesr_cont_fig}
\end{figure}

\begin{figure}[hbt]
\centering
\includegraphics[scale=0.5]{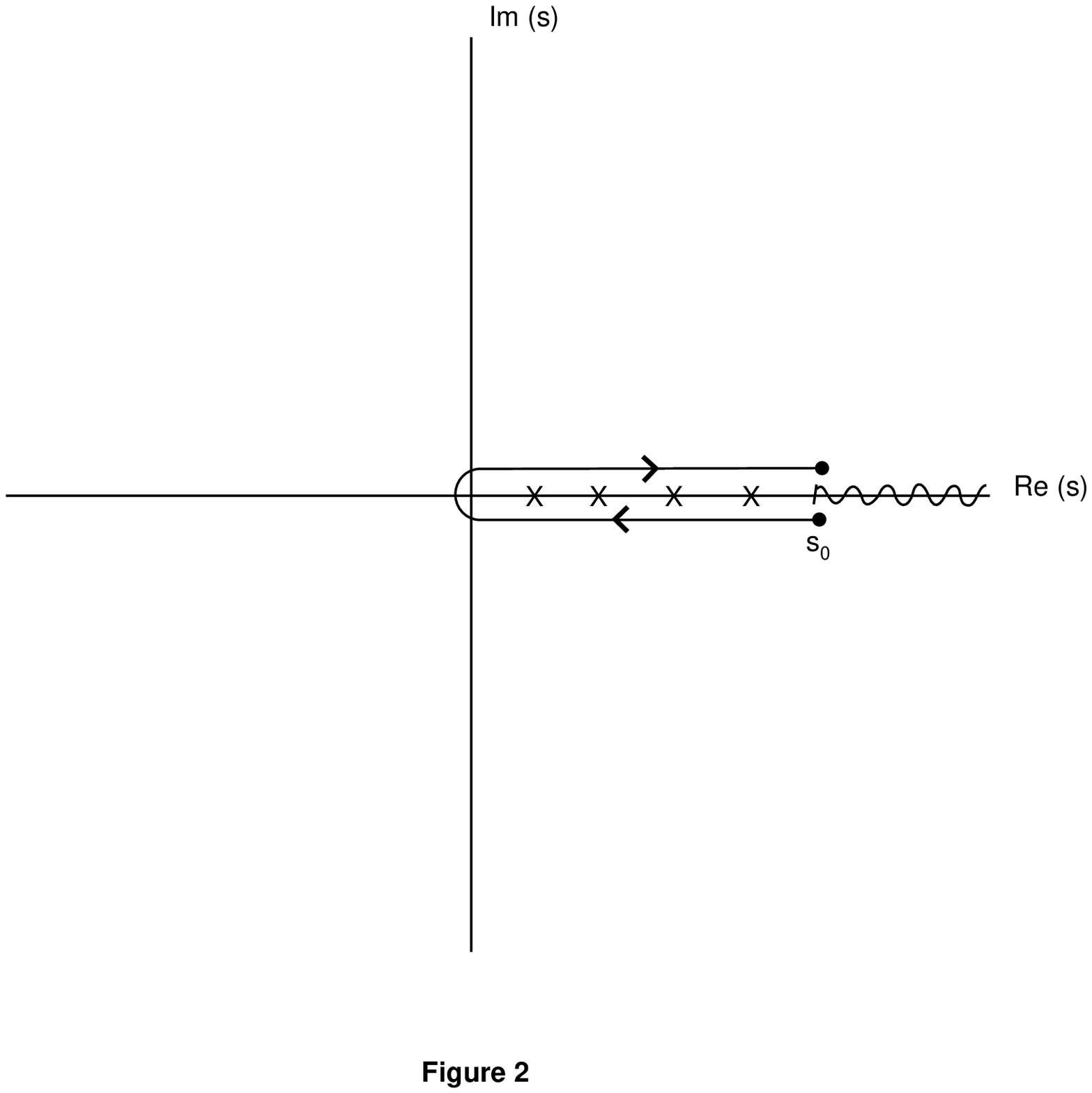}
\caption{
Distortion of the contour $C\left(s_0\right)$ appropriate for resonances and continuum 
contributions on the positive real-$s$ axis. 
}
\label{fesr_cont_fig2}
\end{figure}

The set of FESR's $F_k(s_0) [k \equiv {0,1,2,3,...}]$ is defined by the contour integrals
\begin{equation}
F_k(s_0)\equiv \frac{1}{2\pi i} \,\int_{C(s_0)}\, \Pi(s)s^k ds.
\label{fesr_cont}
\end{equation}
$\Pi(s)$ is, for the case we are considering, the nonstrange scalar-current 
correlation function appropriate for the construction of $\overline{u}u \pm \overline{d}d$
({\it i.e. } $n\bar n$) scalar resonances:
\begin{equation}
\Pi(p^2) = i \,\int\,d^4x\,e^{ip\cdot x} \langle 0 | T J_s (x) J_s (0) |0\rangle,
\label{correlator}
\end{equation}
where, in the SU(2) limit [$m_q \equiv (m_u + m_d)/2$] for isoscalar $(I = 0)$
and isovector $(I = 1)$ currents,
\begin{equation}
J_s(x) = m_q\left[\overline{u}(x)u(x) + (-1)^I\, \overline{d}(x)d(x)\right]/{2} .
\label{current}
\end{equation}
The contour $C(s_0)$ in (\ref{fesr_cont}) is depicted in 
Figure \ref{fesr_cont_fig}.  
Its distortion (Figure \ref{fesr_cont_fig2}) is appropriate for singularities constrained to the positive 
real-$s$ axis, corresponding to resonances and kinematic thresholds for the production of 
physical particles. If the contour radius $s_0$ is chosen to be below these thresholds, 
or if resonances dominate any such thresholds occurring at values of $Re(s)$ below $s_0$, 
one can then model the phenomenological content of (\ref{fesr_cont}) and (\ref{correlator}) by the following 
expressions \cite{Hubschmid}:
\begin{eqnarray}
F_k(s_0) &=& \frac{1}{\pi} \,\int_0^{s_0}\, Im[\Pi(s)]s^k ds,
\label{fesr_phenom}\\
Im[\Pi(s)] &=& 
Im\left[\Pi(s)\right]^{res}
+ \Theta (s - s_0) Im [\Pi^{FT}(s)]
\nonumber\\
&=&
\sum_r \,\pi g_r S_r(s)\Theta \left(s_0 - m_r^2 \right)
+ \Theta (s - s_0) Im [\Pi^{FT}(s)].
\label{res+cont}
\end{eqnarray}
$\Pi^{FT}(s)$ is the  field-theoretical QCD correlation function, which is assumed 
to coincide with its physical (hadronic) counterpart for $s$ sufficiently above the 
resonance region, as characterized by the continuum-threshold parameter $s_0$.  
Each resonance contribution [{\it i.e.}  $Im\Pi^{res}(s)$] to (\ref{res+cont}) is
characterized by 
$S_r(s)$,  a unit-area resonance shape (Breit-Wigner, Gaussian, {\it etc.}) centred at 
the resonance mass $m_r$. The contribution of an individual resonance to the integral of
$\frac{1}{\pi}Im\Pi(s)$ is represented by $g_r$.  
In the narrow-resonance approximation $S_r(s)$ is
\begin{equation}
\lim_{\Gamma_r \rightarrow 0}S_r(s) 
= \delta(s - m_r^2).
\label{narrow}
\end{equation}
To generalize past the narrow resonance approximation, we will assume that $S_r(s)$
is symmetric about $m_r^2$, in which case $S_r(s)$ may be built up as a sum of 
unit-area square pulses centred at $m_r^2$:
\begin{equation}
S_r(s) = \int_0^{\Gamma_{\rm{max}}} \,d\Gamma^\prime \,f(\Gamma^\prime)\left[
\Theta \left(s - m_r^2 + m_r \Gamma^\prime\right) - \Theta 
\left(s - m_r^2 - m_r \Gamma^\prime\right)\right]/2m_r \Gamma^\prime .
\label{res_dist}
\end{equation}
The integrand factor $f(\Gamma^\prime)$ is some unknown width-distribution function, 
such that the full resonance shape has unit area:
\begin{equation}
1 = \int_0^{m_r^2 + m_r \Gamma_{\rm{max}}} \,S_r(s)\,ds .
\label{norm} 
\end{equation}
If we substitute (\ref{res_dist}) and (\ref{res+cont}) into the integrand of (\ref{fesr_phenom}) when $k=0$, 
we find immediately that 
\begin{equation}
F_0(s_0) = \sum_r \, g_r \int_0^{s_0} \,S_r(s)\,ds =
\sum_{m_r^2 < s_0} g_r \quad .
\label{fesr0}
\end{equation}
The final step of (\ref{fesr0}) follows directly from (\ref{norm}) provided all 
resonance peaks are completely below the continuum threshold $s_0$.  
Clearly, the result (\ref{fesr0}) demonstrates that the lowest FESR is impervious to the 
width associated with the shape $S_r(s)$; the same result is obtained by substituting 
the narrow-resonance approximation (\ref{narrow}) into the integrand within the intermediate 
step of (\ref{fesr0}).

Remarkably, this insensitivity  characterizes the second FESR 
$F_1(s_0)$ as well \cite{Deakin}.  To see this, we first substitute (\ref{res+cont}) into (\ref{fesr_phenom}) 
when $k = 1$:
\begin{eqnarray}
F_1(s_0) &=& \sum_r \,\int_0^{s_0} \,S_r(s)\,s \,ds
= \sum_{m_r^2 < s_0} \,g_r \,\int_0^{\Gamma_{\rm{max}}}\,d\Gamma^\prime
\frac{f(\Gamma^\prime)}{2m_r\Gamma^\prime} \,\int_{(m_r^2 - m_r \Gamma^\prime)}
^{(m_r^2 + m_r \Gamma^\prime)}\,ds\, s\nonumber\\
&=& \sum_{m_r^2 < s_0} \,g_r m_r^2 \,\int_0^{\Gamma_{\rm{max}}}\,d\Gamma^\prime\,
f(\Gamma^\prime) .
\label{fesr1_res}
\end{eqnarray}
We see, however, from explicit substitution of (\ref{res_dist}) into (\ref{norm}) that 
\begin{equation}
\int_0^{\Gamma_{\rm{max}}}\,d\Gamma^\prime
f(\Gamma^\prime) = 1, 
\label{dist_norm}
\end{equation}
in which case
\begin{equation}
F_1(s_0) = \sum_{m_r^2 < s_0} \,g_r m_r^2 .
\label{fesr1}
\end{equation}
This result is independent of the width of the (symmetric) resonance peak $S_r(s)$, 
provided the entire resonance peak lies beneath the continuum-threshold $s_0$. 
Indeed, the result (\ref{fesr1}) also characterizes the narrow resonance approximation, 
as is evident upon substitution of (\ref{narrow}) for $S_r(s)$ in the integral appearing 
within the first intermediate step of (\ref{fesr1_res}).

The results (\ref{fesr0}) and (\ref{fesr1}) are useful for constructing phenomenological constraints 
on the lightest $I = 0$ scalar resonance.  The two resonances with which we are 
specifically concerned are $f_0(980)$ and $f_0(400-1200)$; we shall henceforth 
denote the latter resonance as $\sigma$.  Let us first suppose that only one of 
these two low-lying scalar resonances couples strongly to the $I = 0$ $n\bar n$ scalar 
current (\ref{current}).  
One might anticipate such behaviour if $f_0(980)$ were interpreted as a pure $s\bar s$ state, or  
 a non-$q\bar q$ state, such as a $K\overline{K}$ molecule \cite{Weinstein} or a four-quark
$(qq\bar q\bar q)$ exotic \cite{Jaffe}.  As long as $s_0$ is chosen low enough 
to exclude further $I = 0$  $q\overline{q}$ scalar resonances, one finds from (\ref{fesr0}) 
and (\ref{fesr1}) that the lowest lying $q\bar q$ scalar resonance-mass is just
\begin{equation}
[m_\sigma^2]_I = F_1(s_0)/F_0(s_0) .
\label{fesr_ratio1}
\end{equation}
The ``$I$'' subscript implies only a single contributing $I = 0$ scalar resonance 
to the FESR's $F_0$ and $F_1$. Clearly, a $n\bar n$ interpretation of a light 
$\sigma$ is not viable in this scenario if $m_\sigma^2$, as obtained from (\ref{fesr_ratio1}) 
via QCD expressions for $F_{0,1}$, is larger than the experimental $\sim (700 MeV)^2$
upper bound to an empirical light, broad $\sigma$-resonance in $\pi-\pi$ scattering \cite{Sannino,lightsigma}.  
Indeed, if the lowest-lying $n\bar n$ resonance mass were found to be 
near $980 {\rm \,MeV}$, a plausible interpretation would be to identify this resonance 
with $f_0(980)$, and to assume an exotic interpretation for the remaining $f_0(400-1200)$
state.  Indeed, a recent OPAL study of $Z^0$ decays is found to support 
a  $n\bar n$ interpretation of the $f_0(980)$ resonance state \cite{Lafferty}.

The requirement that  $[F_1(s_0)/F_0(s_0)]^{\frac{1}{2}}$ be less than 980 MeV 
for a $n\bar n$ interpretation of a light $\sigma$ is found to hold even 
if we assume there to be {\em two} contributing $m\le 1\,{\rm GeV}$ resonances to scalar current 
$I = 0$ FESR's.  
Suppose both $\sigma$ and $f_0(980)$ resonances contribute to the FESR's $F_{0,1}$.  
Denoting the latter resonance by the subscript $f$, we see from (\ref{fesr0}) and (\ref{fesr1}) that
\begin{eqnarray}
& &F_0 = g_\sigma + g_f ,
\label{f02r}\\
& &F_1 = g_\sigma(m_\sigma^2)_{II} + g_f(m_f^2).
\label{f12r}
\end{eqnarray} 
The subscript ``$II$'' denotes the $\sigma$-mass estimate assuming there to be
two contributing resonances.  One can solve (\ref{f02r},\ref{f12r}) 
for $(m_\sigma^2)_{II}$ to obtain
\begin{equation}
(m_\sigma^2)_{II} = \left[ \frac{F_1(s_0)}{F_0(s_0)} - m_f^2 \right]\,\frac{F_0}{g_\sigma}
+ m_f^2 .
\label{sig_mass_1}
\end{equation}
Since $g_\sigma$, $F_1$, and $F_0$ are all positive, a consequence of the positivity of 
the measure $Im[\Pi(s)] ds$, it is evident from (\ref{sig_mass_1}) that  $(m_\sigma^2)_{II}$
will be smaller than $m_f^2$ [presumably the square of the $f_0(980)$ mass] 
only if $F_1/F_0 < m_f^2$.  Thus, if $m_f$ is identified with the 
$f_0(980)$ mass, we see that $(F_1/F_0)^{\frac{1}{2}}$ must be less than 980 MeV 
for the viability of a $n\bar n$ interpretation for the empirical observation 
of a light $\sigma$.

This result is simplified somewhat by considering the {\em {differences}}  between 
$m_\sigma^2$ and $m_f^2$, where $m_f$ is now to be regarded as {\em{any}} contributing 
second resonance in the $I = 0$ scalar channel:
\begin{eqnarray}
& &\delta_1 \equiv m_f^2 - F_1/F_0 = m_f^2 - (m_\sigma^2)_I , 
\label{delta1}\\
& &\delta_2 \equiv m_f^2 - (m_\sigma^2)_{II} .
\label{delta2}
\end{eqnarray}
We see from (\ref{sig_mass_1}) that 
\begin{equation}
\delta_2 = \frac{F_0}{g_\sigma}\, \delta_1 .
\label{fesr_cons}
\end{equation}
The result (\ref{fesr_cons}) implies that if two resonances contribute to FESR's $F_{0,1}$, it is 
possible to have a very light $\sigma$ coupled to a $n\bar n$
current via a scenario in which 
$\delta_2>\delta_1>0$ [{\it i.e.} 
$\left(m^2_\sigma\right)_{II}<\left(m^2_\sigma\right)_{I}<m_f^2$], provided
 the following two criteria are met:
\begin{enumerate}
\item $g_\sigma$ is small compared to $F_0$, and
\item $(F_1/F_0)^{\frac{1}{2}}$ is smaller than $m_f$, the mass of the heavier resonance.
\end{enumerate}
This latter criterion is necessary for the sign of $\delta_1$ to be positive, 
which is seen from (\ref{fesr_cons}) to be a necessary requirement for positivity of $\delta_2$. 
These criteria generalize to many resonances $m_\sigma<m_1<m_2<\ldots<m_n$ since
\begin{equation}
\frac{F_1}{F_0}=\frac{g_\sigma m_\sigma^2+g_1m_1^2+g_2m_2^2+\ldots +g_nm_n^2}{g_\sigma+g_1+g_2+\ldots +g_n}
\le m_n^2\quad .
\end{equation}
Thus the second criterion above should be understood as a constraint on the {\em heaviest} contributing resonance 
consistent with the choice of continuum threshold $s_0$.  
Although the first  criterion listed above is quite \\
reasonable,
\footnote{It has been argued
elsewhere \cite{first_paper,Clement} 
that $g_\sigma \sim (m_q^2/4)f_\pi^2 m_\pi^2$, where we have included
an additional factor of $(m_q^2/4)$ as a consequence of the additional factor of
$(m_q/2)$ characterizing the scalar current (\protect\ref{current}) relative to that of ref. \cite{Clement}.}
the latter criterion provides an important constraint on any attempt to model an 
empirical light broad $\sigma$ as a $q\overline{q}$ state.  

\section{Field-Theoretical Content of Scalar-Current FESR's}
\label{fesr_theor_sec}
\begin{figure}[hbt]
\centering
\includegraphics[scale=0.5]{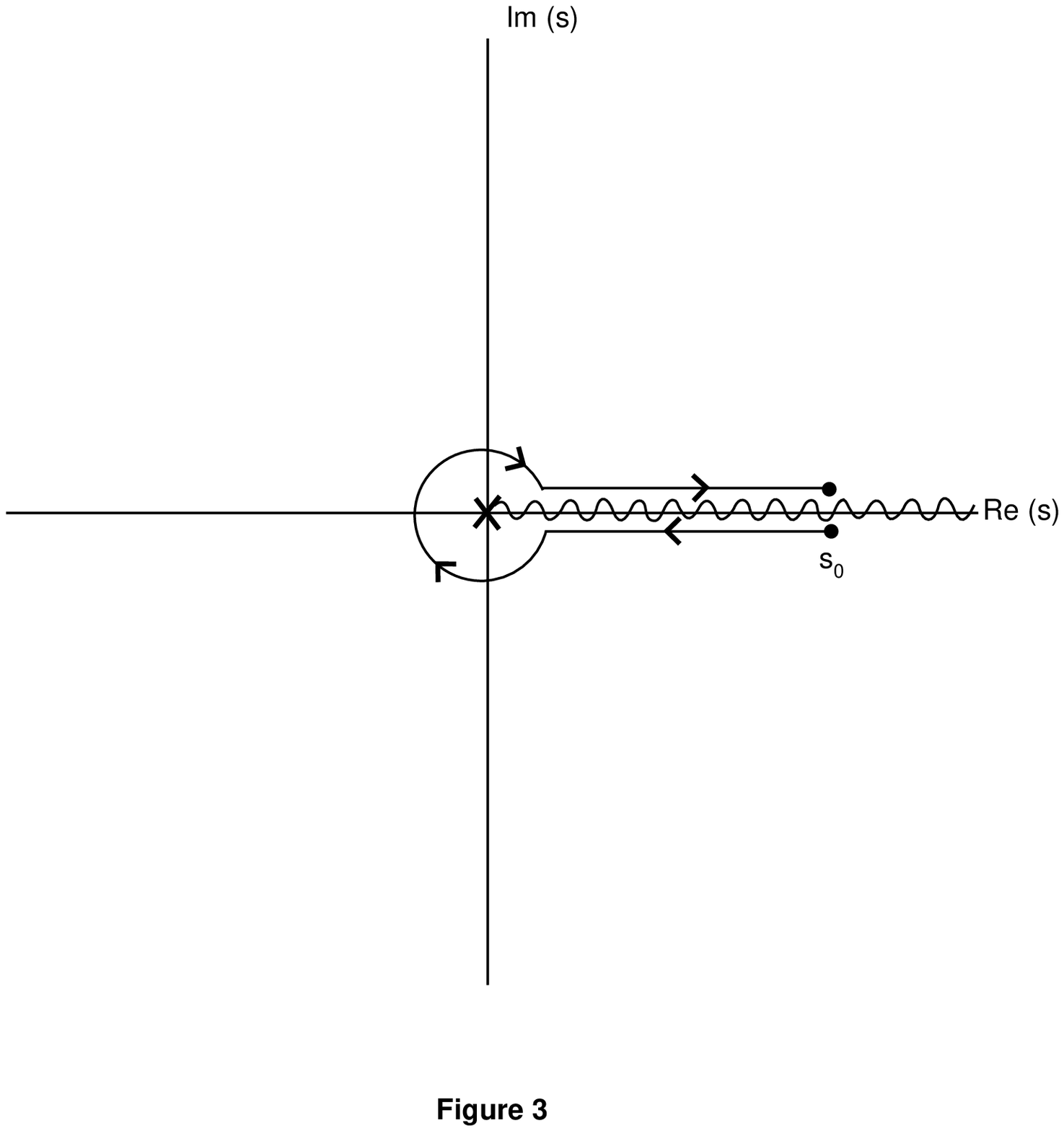}
\caption{
Distortion of the contour $C\left(s_0\right)$ appropriate
for field-theoretical poles at $s=0$ and field-theoretical branch singularities along the 
entire positive real-$s$ axis.
}
\label{fesr_cont_fig3}
\end{figure}

QCD field-theoretical expressions for the sum rules $F_{0,1}(s_0)$ are obtained 
from purely-perturbative, vacuum condensate, and instanton contributions to the 
scalar correlator $\Pi(s)$, as defined by (\ref{correlator}).  
Both the purely-perturbative and the instanton contributions
(nonperturbative contributions of finite correlation length) 
can be understood to have a branch singularity along the positive real-$s$ axis of Figure 
\ref{fesr_cont_fig}.  
By contrast, the QCD-vacuum condensate contributions (nonperturbative contributions 
of infinite correlation length) are characterized by poles at $s = 0$, 
as obtained from the Wilson coefficients within the operator product expansion of 
the correlator.  Consequently, purely-perturbative (pert), vacuum condensate (cond), 
and instanton (inst) contributions to the FESR's $F_k(s_0)$ can be obtained from 
(\ref{fesr_cont}) by distorting the contour $C(s_0)$ as indicated in Figure \ref{fesr_cont_fig3}:
\begin{equation}
F_k(s_0) = \frac{1}{\pi} \,\int_0^{s_0} \left\lbrace Im \left[\Pi(s, \mu^2 = s_0)
\right]_{\rm{pert}} + Im[\Pi(s)]_{\rm{inst}} \right\rbrace s^k ds-
{\rm{Res}}_{s=0}
\left\lbrace s^k[\Pi(s)]_{\rm{cond}} \right\rbrace .
\label{fesr_theor}
\end{equation}
The leading condensate contributions to the scalar correlation function 
can be extracted from refs. \cite{Bagan,Reinders,SVZ}:
\begin{eqnarray}
[\Pi(s = -Q^2)]_{\rm{cond}} &=& \frac{3m_q^2}{2Q^2} \langle m_q \overline{q}q\rangle +
\frac{m_q^2}{16\pi Q^2} \langle\alpha_s G^2 \rangle
+\frac{m_q^2 \pi}{Q^4}\langle{\cal O}_6\rangle
+\frac{m_q^3}{2Q^4} \langle\overline{q} G\cdot \sigma q\rangle
\nonumber\\
& &+ m_q^4\left[C_1\,\frac{\langle m_q\overline{q}q\rangle}{Q^4} +
C_2 \,\frac{\langle\alpha_s G^2 \rangle}{Q^4}
 + C_3 \,\frac{\langle\alpha_s (\overline{q}q)^2\rangle}{Q^6} +
C_4 \,\frac{\langle g_s G^3\rangle}{Q^6} \right] . 
\label{condensates}
\end{eqnarray}
The quantity  $\langle {\cal O}_6\rangle$ denotes the following linear combination of dimension-six quark condensates:
\begin{equation}
\langle{\cal O}_6\rangle\equiv \alpha_s
\left[ 
\frac{1}{4}\langle \left(\bar u\sigma_{\mu\nu}\lambda^a u-\bar d\sigma_{\mu\nu}\lambda^a d\right)^2\rangle
+\frac{1}{6}
\langle \left(   
\bar u \gamma_\mu \lambda^a u+\bar d \gamma_\mu \lambda^a d 
\right)
\sum_{u,d,s}\bar q \gamma^\mu \lambda^a q
\rangle\right]\quad .
\label{o6}
\end{equation} 
The  vacuum saturation hypothesis \cite{SVZ} in the SU(2) limit
$\langle \bar u u\rangle=\langle\bar d d\rangle\equiv\langle\bar q q\rangle$ provides  a reference value
for $\langle {\cal O}_6\rangle$
\begin{equation}
\langle{\cal O}_6\rangle=-f_{vs}\frac{88}{27}\alpha_s
\langle (\bar q q)^2\rangle
=-f_{vs}5.9\times 10^{-4} {\rm GeV}^6\quad , 
\label{o61}
\end{equation}
where $f_{vs}=1$ for exact vacuum saturation. The quark condensate is determined by the GMOR (PCAC)
relation \cite{GellMann}, and the gluon condensate is given by \cite{dimsix}
\begin{equation}
\langle \alpha_s G^2\rangle=\left(0.045\pm0.014\right) \,{\rm GeV^4}\quad .
\end{equation}
The coefficients $C_1 - C_4$ of $O(m_q^4)$ terms are dimensionless 
and independent of $m_q$ after operator re-alignment \cite{Jamin}  is imposed to circumvent 
mass singularities [$C_2$ and $C_4$ are seen \cite{Deakin}
to still include factors of $\ln(Q^2/\mu^2)$].  
We see that all but the first three terms on the right hand side of (\ref{condensates}) are chirally 
suppressed; the condensate $\langle m_q\overline{q}q\rangle = -f_\pi^2 m_\pi^2/2$ \cite{SVZ,GellMann}
is understood 
to have no explicit $m_q$-dependence ($f_\pi \equiv 93 {\rm MeV}$). 
These three leading terms contribute to the residue portion of (\ref{fesr_theor}).

The $n_f = 3$ purely-perturbative QCD contribution to the imaginary part of the 
scalar correlator (\ref{correlator}) has been calculated to four-loop order \cite{Chetyrkin}:
\begin{eqnarray}
\frac{1}{\pi} \,Im \left[\Pi(s,\mu^2)\right]_{\rm{pert}}
&=& \frac{3m_q^2 s}{16\pi^2} \Biggl[1 + \frac{\alpha_s}{\pi}
\left(\frac{17}{3} - 2 \ln
\left(\frac{s}{\mu^2}\right)\right)
\Biggr.\nonumber\\
& &\qquad+ \left(\frac{\alpha_s}{\pi}\right)^2 \left(31.8640 - \frac{95}{3} \ln
\left(\frac{s}{\mu^2}\right) + \frac{17}{4} \ln^2 
\left(\frac{s}{\mu^2}\right)
\right)
\label{ImPi}
\\
& &\qquad+ \Biggl.\left(\frac{\alpha_s}{\pi}\right)^3 \left(89.1564 - 297.596 \ln
\left(\frac{s}{\mu^2}\right) + \frac{229}{2} \ln^2 
\left(\frac{s}{\mu^2}\right) - \frac{221}{24} \ln^3 \left(\frac{s}{\mu^2}\right)
\right)\Biggr] ~.
\nonumber
\end{eqnarray}
The renormalization-group (RG) invariance of the correlator (\ref{correlator}), 
as obtained from the RG invariant scalar current (\ref{current}), allows RG improvement 
within the sum rule integrand (\ref{ImPi}) as follows ($\mu^2 \equiv s_0$):
\begin{eqnarray}
& &\alpha_s \longrightarrow \alpha_s(\sqrt{s_0}),
\label{rg1}\\
& &m_q \longrightarrow m_q(\sqrt{s_0}),
\label{rg2}
\end{eqnarray}
with the evolution of $\alpha_s(\mu)$ and $m_q(\mu)$ given by four-loop
$\beta$ and $\gamma_m$ functions derived in refs. \cite{beta}
and \cite{ChetyrkinRitbergen}, respectively (see
\cite{run_alpha,run_mass}  for explicit forms).

The imaginary part of the direct single-instanton contribution to the scalar correlator 
in the dilute instanton-liquid model is given by \cite{instanton_continuum,Deakinetal}
\begin{equation}
\frac{1}{\pi} \,Im\left[\Pi(s)\right]_{\rm{inst}} = -\frac{3m_q^2
s}{8\pi}\, J_1 (\rho\sqrt{s})Y_1(\rho\sqrt{s}), 
\label{ImPi_ins}
\end{equation}
where $\rho$ is the (uniform) instanton size $\rho=1/(600\, {\rm MeV})$ appropriate 
for the instanton-liquid model \cite{ins_liquid}.  We can now substitute (\ref{condensates}--\ref{ImPi_ins}) 
into (\ref{fesr_theor}) 
to obtain explicit expressions for the first two FESR's:
\begin{eqnarray}
\frac{F_0(s_0)}{ m_q^2 (s_0)}& =&  
 \frac{3s_0^2}{32\pi^2} \left[1 + \frac{20}{3\pi} \alpha_s(\sqrt{s_0}) +
\frac{49.8223}{\pi^2} \, \alpha_s^2 (s_0) + \frac{302.110}{\pi^3}
\alpha_s^3(\sqrt{s_0}) +
 {\cal{O}} (\alpha_s^4)
\right]
\nonumber\\
& &
+\frac{3}{2} \langle m_q \overline{q} q\rangle +
\frac{1}{16\pi} \langle\alpha_s G^2\rangle + {\cal{O}} (m_q^2) 
- \frac{s_0^2}{8\pi} \left[J_1(\rho\sqrt{s_0})Y_1(\rho\sqrt{s_0})
 + J_2 (\rho\sqrt{s_0})Y_2 
(\rho\sqrt{s_0})\right]\quad ,
\label{F0}
\end{eqnarray}
\begin{eqnarray}
\frac{F_1(s_0)}{ m_q^2 (s_0)} &=&
 \frac{s_0^3}{16\pi^2} \left[1 + \frac{19}{3\pi} \alpha_s (\sqrt{s_0}) +
\frac{43.3640}{\pi^2} \, \alpha_s^2 (s_0) + \frac{215.846}{\pi^3}
\alpha_s^3 (\sqrt{s_0}) 
+  {\cal{O}} (\alpha_s^4)
\right]
\nonumber\\
& & - \frac{3s_0^3}{160\pi} \left[5J_1 (\rho\sqrt{s_0})Y_1
(\rho\sqrt{s_0})+ 4J_2(\rho\sqrt{s_0})Y_2(\rho\sqrt{s_0})
- J_3(\rho\sqrt{s_0})Y_3(\rho\sqrt{s_0})\right] 
\nonumber\\
& &+\frac{88\pi}{27}
f_{vs}\langle\alpha_s (\overline{q}q)^2\rangle + {\cal{O}}(m_q)
\quad .
\label{F1}
\end{eqnarray}
We reiterate that these FESR's are dual to phenomenological expressions that 
are independent of resonance-shape effects, as long as such resonance shapes 
are symmetric and entirely beneath the continuum threshold $s_0$.  The instanton contributions 
to (\ref{F0}) and (\ref{F1}) are respectively obtained via the following indefinite integrals \cite{Xue}:
\begin{eqnarray}
& &\int z^3 J_1(z) Y_1(z)dz = \frac{z^4}{6} \left[ J_1 (z) Y_1 (z) +
J_2(z)Y_2(z)\right] + {\rm constant},\\ 
& &\int z^5 J_1(z) Y_1(z)dz = \frac{z^6}{8} \left[ J_1 (z) Y_1 (z) +
\frac{4}{5} J_2(z) Y_2 (z)
- \frac{1}{5} J_3(z)Y_3(z)\right] + {\rm constant}~~.
\end{eqnarray}

\begin{figure}[hbt]
\centering
\includegraphics[scale=0.7]{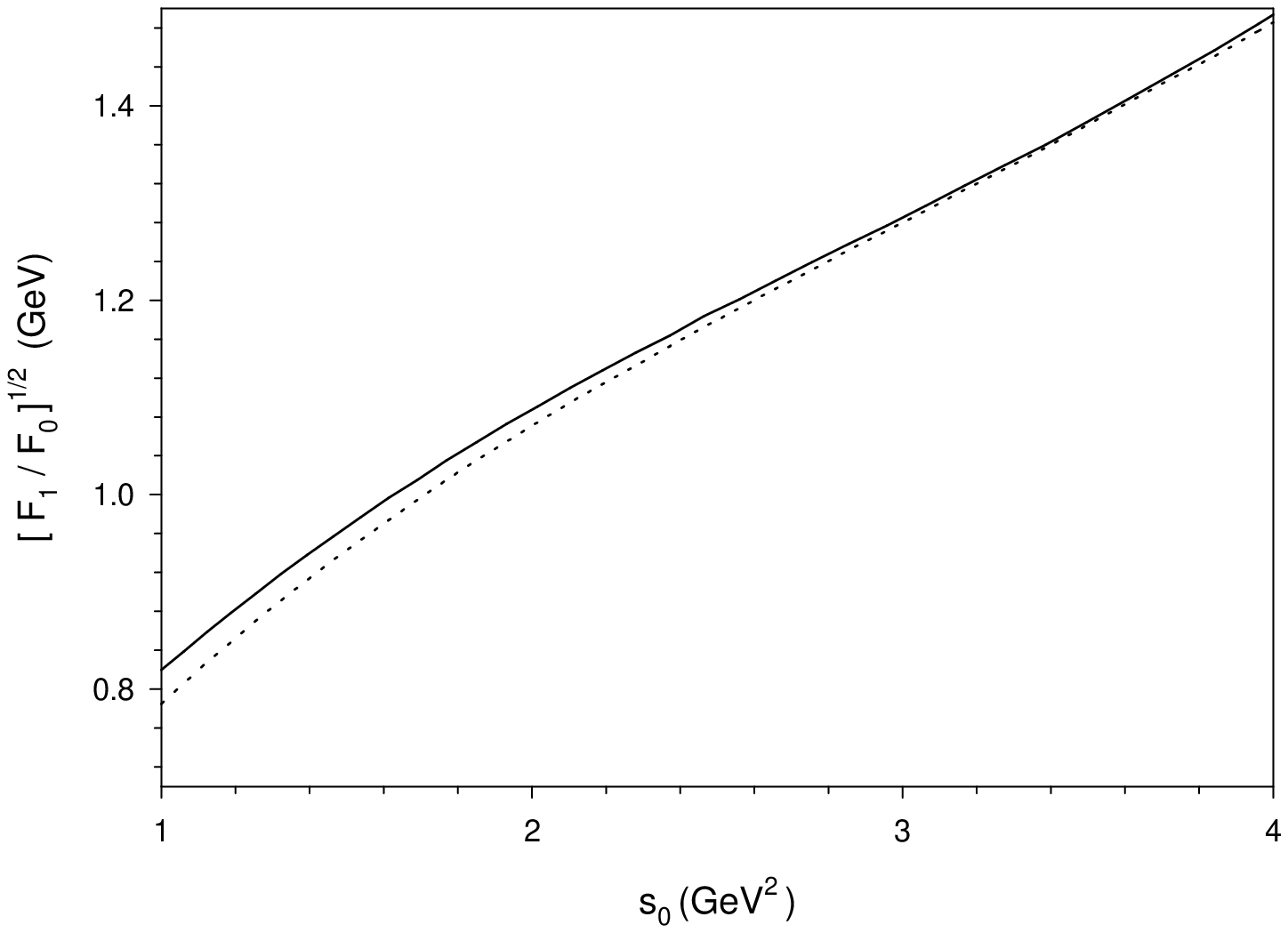}
\caption{
The quantity $\sqrt{F_1/F_0}$ is plotted as a function of $s_0$ using central values of the nonperturbative 
parameters as outlined in the text.  The solid curve uses two-loop perturbative corrections, while the dotted curve uses
four-loop corrections. 
}
\label{fesr_ratio_fig}
\end{figure}

\begin{figure}[hbt]
\centering
\includegraphics[scale=0.7]{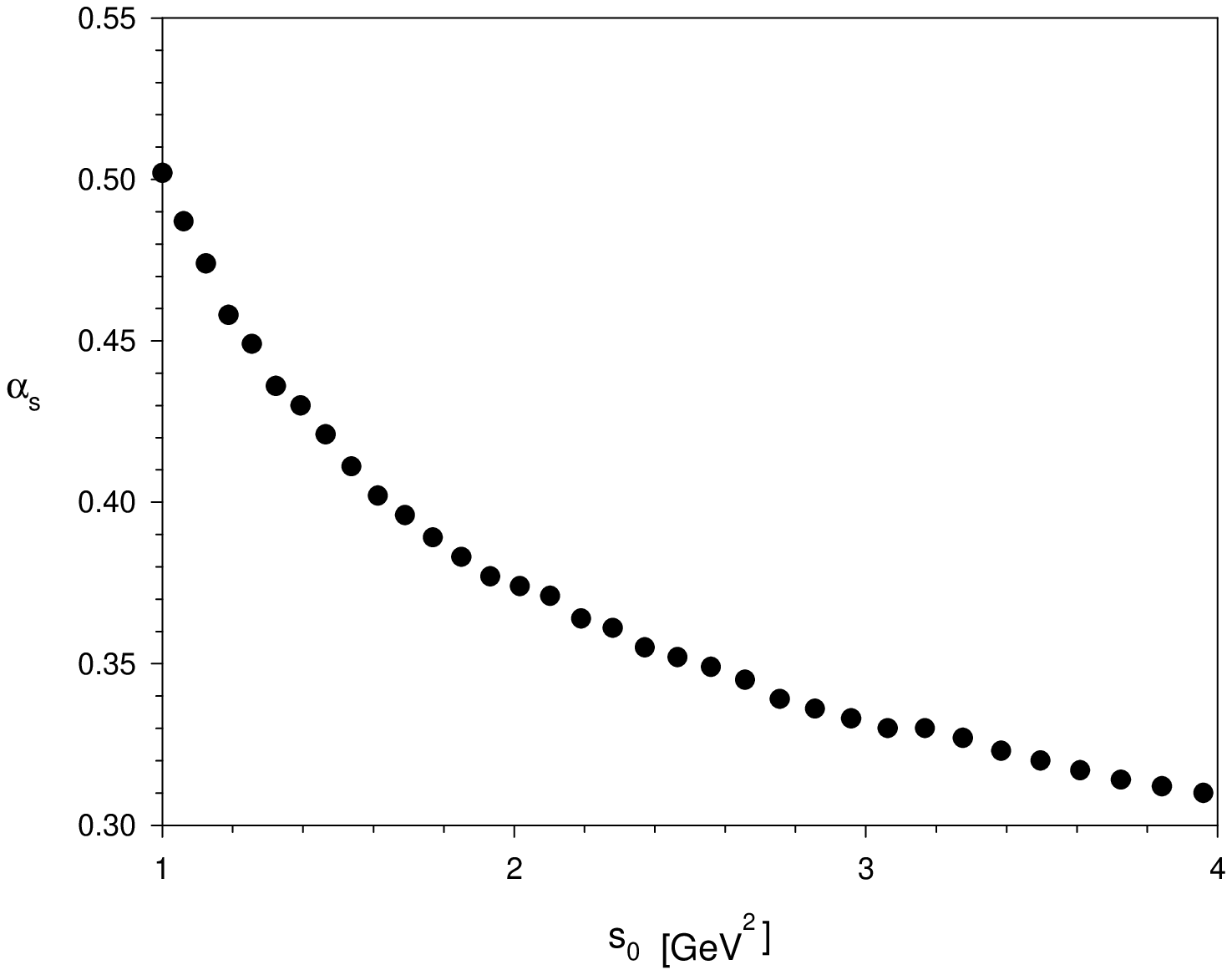}
\caption{
Evolution of the running coupling $\alpha_s$ versus the energy scale $s_0$
using $\alpha_s\left(M_Z\right)$ as an initial condition.  Matching conditions
imposed at the flavour thresholds are evident. 
}
\label{coupling_fig}
\end{figure}

Equations (\ref{F0}) and (\ref{F1}) are sufficient to generate via (\ref{fesr_ratio1}) the estimate
$(m_\sigma)_I$ of the lowest-lying $I = 0$ scalar meson, assuming this to be the {\em{only}} 
contributing resonance. In Figure \ref{fesr_ratio_fig}, $(m_\sigma)_I [\equiv (F_1/F_0)^{\frac{1}{2}}]$ 
is displayed as a function of the continuum threshold parameter $s_0$ up to values 
comparable to $s_0=4\,{\rm GeV^2}$, using the central 
nonperturbative-parameter values defined above. The $s_0$-dependence 
of the coupling-constant $\alpha_s(\sqrt{s_0})$, which explicitly enters the right-hand sides 
of (\ref{F0}) and (\ref{F1}), is displayed in Figure \ref{coupling_fig}, and is obtained 
(as described in ref. \cite{pade}) 
from the initial condition $\alpha_s(M_Z) = 0.119$ via the four-loop $\beta$-function 
with 4- and 5-flavour threshold discontinuities \cite{ChetyrkinKniehl} at 1.3 GeV and 4.3 GeV, 
respectively.\footnote{For the four-loop curve in Figure \protect\ref{fesr_ratio_fig}, we have used $n_f=4$ expressions
obtained from \protect\cite{Chetyrkin} for the perturbative content of $F_{0,1}$ when $\sqrt{s_0}> 1.3\,{\rm GeV}$,
the four-flavour threshold.  The 2-loop expressions for $F_{0,1}$ are independent of $n_f$ \protect\cite{Chetyrkin}.  } 
From Figure \ref{fesr_ratio_fig}, we see that $(m_\sigma)_I$ increases with $s_0$ and is {\em{never}} comparable 
to the empirical 400-700 MeV mass range anticipated from $\pi-\pi$ scattering phase shifts \cite{Sannino,lightsigma}; 
even for the (unrealistically low) choice of  $1\,{\rm GeV}^2$ for the continuum-threshold, 
the single-resonance fit for $m_\sigma$ is nearly  $800\, {\rm MeV}$.

 We see from Figure \ref{fesr_ratio_fig} that for 
values  $1.0\,{\rm GeV}\le \sqrt{s_0}\le 1.3\,{\rm GeV}$ for which the  $f_0(980)$ peak 
(but no subsequent resonance) is expected to contribute, the range of the single-resonance 
estimate of the lowest-lying $I = 0$ $n\bar n$-resonance appears consistent with that 
resonance actually being $f_0(980)$.  Such an interpretation of this state is consistent 
with a recent analysis of OPAL data \cite{Lafferty}, but would necessarily require a 
non-$n\bar n$
interpretation (or a mass near or above 1 GeV) for the $f_0(400-1200)$ state. 

A drawback to the finite energy sum rule approach is its failure to suppress 
non-low-lying resonances, as is evident from (\ref{fesr0}) and (\ref{fesr1}). 
Recall that in the previous section, we examined the phenomenological consequences of having multiple 
resonances contribute to the FESR's (\ref{F0}) and (\ref{F1}).  A light resonance is found to be 
possible only if  $(F_1/F_0)^{\frac{1}{2}}$ is smaller than the mass of the heaviest 
contributing resonance.  In a scenario in which there are now two contributing $n\bar n$
resonances, the $f_0(980)$ and a lighter (400-700 MeV) $\sigma$, we see  from 
Figure \ref{fesr_ratio_fig} that the continuum threshold parameter $s_0$ is restricted to values 
below $1.6 \,{\rm GeV^2}$, the value of $s_0$ at which $(F_1/F_0)^{\frac{1}{2}}$ is equal 
to 980 MeV. In Section \ref{holder_sec}, we shall demonstrate via Holder-inequalities that this range 
of $s_0$ is excluded from sum-rule parameter space -- consequently, 
we can rule out the scenario in which $f_0(980)$ and an even lighter $\sigma$ 
{\em{both}} contribute (as $n\bar n$-resonances) to the first two FESR's.  
We cannot, however, exclude the possibility of such a light $\sigma$ hiding under the large 
contribution of $f_0(1370)$, assuming an $n\bar n$ interpretation for this state \cite{Spanier}, 
as the value of $(F_1/F_0)^{\frac{1}{2}}$ is still below $1370 \,{\rm MeV}$ for values of $s_0$ 
as large as $3.5 \,{\rm GeV^2}$. Indeed,  similar  behaviour characterizes the pseudoscalar 
channel, where sum rule contributions from the first pion excitation [$\Pi(1300)$] 
dominate those of the lowest-lying pseudoscalar resonance, the pion itself \cite{Steele}.   

Summarizing, we find that the first two finite-energy sum rules have the advantage of 
being insensitive to widths of (symmetric) subcontinuum resonance peaks. They have the disadvantage of failing to 
suppress contributions from non-lowest-lying subcontinuum resonances.  For the suppression of such 
resonances within a sum rule context, we necessarily must employ width-sensitive Laplace 
sum rules, as previously considered in ref. \cite{first_paper}. Such sum rules $R_k(\tau,s_0)$ are 
defined similarly to the FESR's (\ref{fesr_cont}) except for the occurrence of an exponential 
damping factor $e^{-s\tau}$ in the sum-rule integrand: 
\begin{equation}
R_k(\tau, s_0) = \frac{1}{2\pi i} \,\int_{C(s_0)}\,\Pi(s) s^k e^{-
s\tau}ds.
\label{laplace}
\end{equation}
If we substitute the phenomenological resonance content of $Im\Pi$, as given in 
(\ref{res+cont}), and utilize the narrow resonance approximation (\ref{narrow}), we find there to be a 
progressive exponential suppression of heavy resonances within such sum rules \cite{SVZ}, 
in contrast to the resonance content of corresponding FESR's given by (\ref{fesr0}) and (\ref{fesr1}): 
\begin{eqnarray}
& &R_0(\tau,s_0) = \sum_{m_r^2 < s_0} \,g_r e^{-m_r^2 \tau},
\label{phenom_lsr0}\\
& &R_1(\tau,s_0) = \sum_{m_r^2 < s_0} \,g_r m_r^2 e^{-m_r^2 \tau}.
\label{phenom_lsr1}
\end{eqnarray}
The contribution of the lowest-lying resonance $(\ell)$ is pronounced provided the Borel-parameter 
$\tau$ is chosen so as to be less than the reciprocal of the square of the mass of the 
lightest resonance $(m_\ell^2 \tau < 1)$, but greater than the reciprocal of the square of 
the mass of any subsequent $(h)$ resonance $(m_h^2 \tau > 1)$.  In the section which follows, 
we shall examine in detail the field theoretical content of the leading Laplace 
sum rule in the scalar channel, particularly its instanton content as extracted from (\ref{laplace}).

\section{Field-Theoretical Content of the Laplace Sum-Rule}
\label{laplace_theor_sec}
The leading Laplace sum-rule $R_0(\tau,s_0)$ may be obtained from (\ref{laplace}) by distorting the contour
$C(s_0)$ as indicated in Figure \ref{fesr_cont_fig3}, a procedure identical to that by which the FESR's
(\ref{fesr_theor}) are calculated from (\ref{fesr_cont}):  
\begin{equation}
R_0(\tau,s_0)=\frac{1}{\pi}\int\limits_0^{s_0}\left\lbrace
Im\left[\Pi(s)\right]_{pert}+Im\left[\Pi(s)\right]_{inst}
\right\rbrace e^{-s\tau} ds
-{\rm Res}_{s=0}\left\lbrace e^{-s\tau} \left[\Pi(s)\right]_{cond}   \right\rbrace\quad .
\label{laplace_theor}
\end{equation}
It is customary, however, to express this sum-rule as the difference between its $s_0\rightarrow\infty$ limit
and a continuum-contribution $c_0(\tau,s_0)$, which is seen from (\ref{laplace_theor}) to be the following:
\begin{eqnarray}
& &R_0(\tau,s_0)=R_0(\tau,\infty)-c_0(\tau,s_0)\quad ,
\label{sr_cont}\\
& &R_0(\tau,s_0)
=\frac{1}{\pi}\int\limits_0^{s_0} Im\Pi^{res}(s) e^{-s\tau}\,ds \quad , 
\label{res_sr}
\\
& &c_0(\tau,s_0)=\frac{1}{\pi}\int\limits_{s_0}^\infty\left\lbrace
Im\left[\Pi(s)\right]_{pert}+Im\left[\Pi(s)\right]_{inst}
\right\rbrace e^{-s\tau} ds=c_0^{pert}(\tau,s_0)+c_0^{inst}(\tau,s_0) \quad .
\label{c0}
\end{eqnarray}
We see from substitution of (\ref{condensates}), (\ref{ImPi}), and (\ref{ImPi_ins}) into the $s_0\rightarrow\infty$
limit of (\ref{laplace_theor}) that, for $I=\{0,1\}$ \cite{Reinders,SVZ,Dorokhov},
\begin{eqnarray}
{ R}_0(\tau,\infty)&=&\frac{3m_q^2}{16\pi^2\tau^2}\left(   
1+4.821098 \frac{\alpha_s}{\pi}
+{\cal O}\left[\alpha_s^2\right]\right)
+m_q^2\left(
\frac{3}{2}\langle m_q\bar q q\rangle 
+\frac{1}{16\pi}\langle \alpha_s G^2\rangle
+\pi\langle{\cal O}_6\rangle \tau
\right)
\nonumber
\\
& & \qquad +\left(-1\right)^I m_q^2
{3\rho^2\over{16 \pi^2\tau^3}} e^{-\frac{\rho^2}{2\tau} }
\left[   
  K_0\left( \frac{\rho^2}{2\tau} \right) +
       K_1\left(\frac{\rho^2}{2\tau}  \right)
\right]\quad .
\label{sr_nc}
\end{eqnarray}
We have included within (\ref{sr_nc}) only the two-loop perturbative contribution, as
condensate contributions are known to only one-loop order.  Sum-rule stability under 
higher-loop perturbative expressions will be discussed in section \ref{hl_sec}.  The sum-rules $R_k(\tau,\infty)$ are known 
\cite{Narison} to satisfy RG equations with respect to the mass scale $\mu=1/\sqrt{\tau}$, thereby justifying the 
following two-loop order RG improvements:
\begin{eqnarray}
& &\frac{\alpha_s}{\pi}\rightarrow \frac{\alpha_s(\mu)}{\pi}= \frac{1}{\beta_0 L}-\frac{\beta_1\log L}{\beta_0\left(\beta_0L\right)^2}
\quad ,
\label{alpha}\\
& &L=\log\left(\frac{\mu^2}{\Lambda^2}\right)\quad ,\quad 
\beta_0=\frac{9}{4}\quad ,\quad \beta_1=4\quad ;
\\
& &m_q\rightarrow m_q(\mu)\equiv \hat m_q G(\mu) \quad ,
\label{hatm}
\\
& &G(\mu)=\frac{1}{\left(\frac{1}{2}L\right)^{\frac{4}{9}}}\left(
1+\frac{290}{729}\frac{1}{L}-\frac{256}{729}\frac{\log{ L}}{L}
\right) \quad .
\end{eqnarray}
We use $\Lambda_{\overline{MS}}\approx 300\,{\rm MeV}$ for three active flavours,
consistent with current estimates of $\alpha_s(M_\tau)$ \cite{PDG,pade_tau} and matching conditions through the charm threshold 
\cite{ChetyrkinKniehl}.

The above formulation differs from past treatments (even those in which the instanton contribution to $R_0$ is explicit 
\cite{first_paper}) in which only purely-perturbative effects contribute to the continuum (\ref{c0}).  We see
from substitution of (\ref{ImPi_ins}) into (\ref{c0}) that methodological consistency requires an explicit instanton 
 contribution to the continuum,
\begin{equation}
c_0^{inst}\left(\tau,s_0\right)
=
\left(-1\right)^{I+1}\frac{3m_q^2}{8\pi}
\int\limits_{s_0}^\infty
s J_1\left(\rho\sqrt{s}\right)Y_1\left(\rho\sqrt{s}\right)e^{-s\tau}\,ds\quad ,
\label{ins_continuum}
\end{equation}
in addition to the known perturbative contribution arising from (\ref{ImPi}):
\begin{eqnarray}
& &c_0^{pert}\left(\tau,s_0\right)
= \frac{3m_q^2}{16\pi^2}\left[\left(1+\frac{17}{3}\frac{\alpha_s}{\pi}\right)
f_0\left(\tau, s_0\right)
-2\frac{\alpha_s}{\pi}
f_1\left(\tau, s_0\right)
\right] \quad ,
\label{pert_cont}
\\
& & f_0\left(\tau, s_0\right)=\frac{1}{\tau^2}\left(1+s_0\tau\right)e^{-s_0\tau}\quad ,
\label{f0}\\
& &f_1\left(\tau,s_0\right)=\frac{1}{\tau^2}\left[
\left(1+s_0\tau\right)e^{-s_0\tau}\log\left(s_0\tau\right)+e^{-s_0\tau}+E_1\left(s_0\tau\right)
\right]\quad .
\label{f1}
\end{eqnarray}
In the above equations $\gamma_E$ is Euler's constant and $E_1(x)$ is the exponential integral.

\begin{figure}[hbt]
\centering
\includegraphics[scale=0.7]{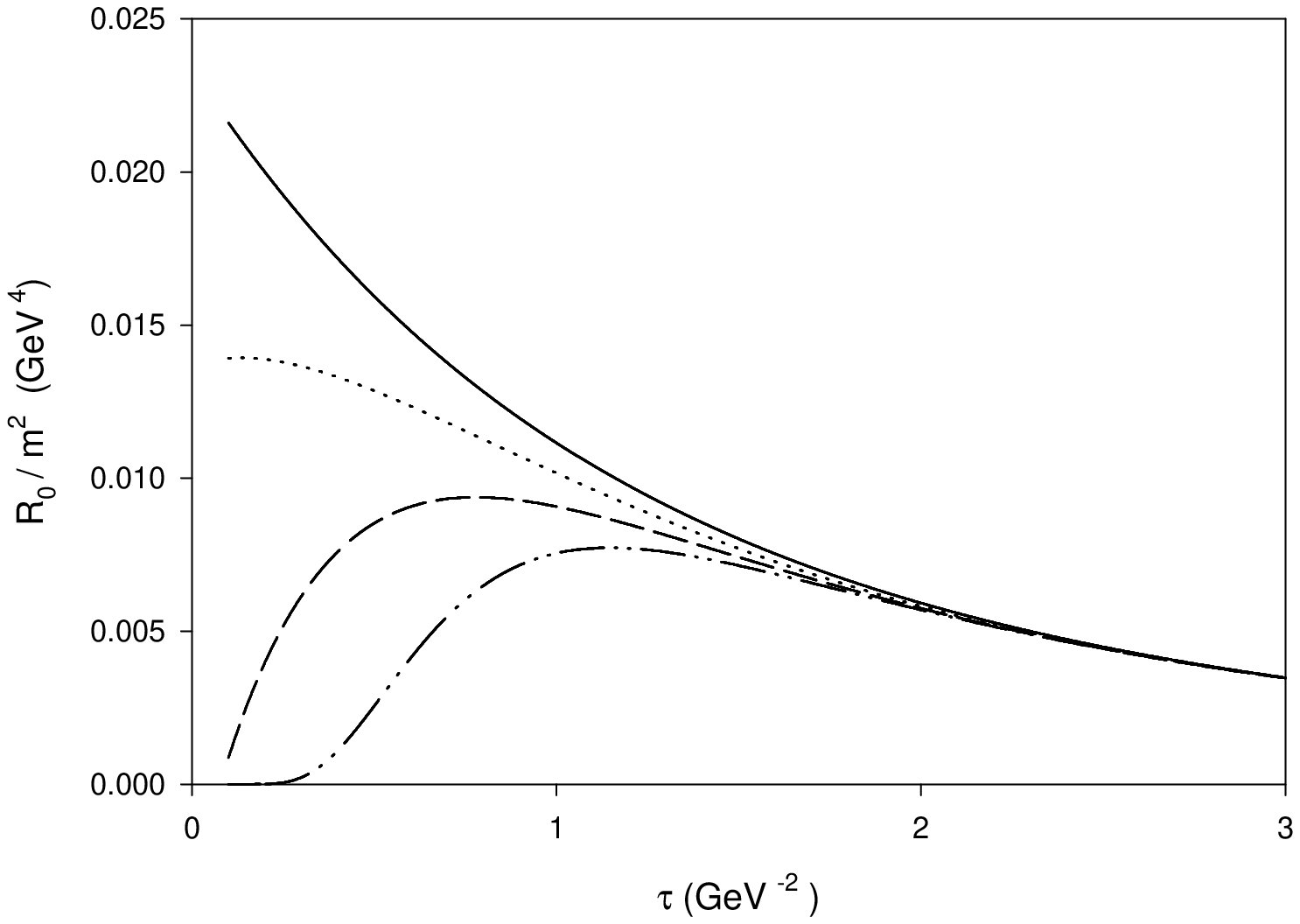}
\caption{
Comparison of the instanton contributions to the Laplace sum-rule with 
inclusion of the 
instanton continuum (upper three curves), and omission of the instanton continuum
(lowest dashed-dotted curve).  The $s_0$ values used for evaluation
of the  instanton continuum are: $s_0=2.5\,{\rm GeV^2}$ (solid curve),
$s_0=3.0\,{\rm GeV^2}$ (dotted  curve), and
$s_0=3.5\,{\rm GeV^2}$ (dashed curve).
}
\label{ins_fig}
\end{figure}

We reiterate that the instanton continuum contribution has been ignored in previous applications 
of instanton effects in sum-rules.  
To determine the numerical significance of this contribution, we compare the total instanton contributions to the sum-rule
(\ref{laplace_theor}) before and after inclusion of the instanton continuum.  As shown in Figure \ref{ins_fig},
inclusion of the continuum {\em enhances} the total instanton contributions.  Since the instanton effects without the continuum
are responsible for the $\sim 500\,{\rm MeV}$ splitting between the lowest-lying 
$I=0$ and $I=1$ states found in \cite{first_paper}, enhancement of the instanton effects suggests that an even larger isospin 
splitting could occur by including the instanton contribution to the continuum (\ref{c0}).
Furthermore, it is interesting to note that integration of (\ref{ins_continuum}) 
with the instanton density $d(\rho)$ will be well behaved at small $\rho$ because  $J_1\left(\rho\sqrt{t}\right)$ goes to zero 
for small $\rho$.  This behaviour should be contrasted with that of \cite{instanton_continuum,SVZ} 
\begin{equation}
\Pi^{inst}(s=-Q^2)=\left(-1\right)^I\frac{3}{4\pi^2}m_q^2Q^2
\left[K_{-1}\left(\rho\sqrt{Q^2}\right)\right]^2 \quad ,
\label{instanton}
\end{equation}
which   has  an infrared divergence when integrated over $\rho$.

\section{H\"older Inequality Constraints on the Laplace Scalar Sum-Rules}
\label{holder_sec}
In the phenomenological analysis of QCD sum-rules, the $\tau$-dependence of $R_0(\tau,s_0)$,
as given by (\ref{sr_cont}), (\ref{sr_nc}), (\ref{ins_continuum}) and (\ref{pert_cont}), is used to extract phenomenological resonance parameters $g_r$ and 
$m_r^2$ within (\ref{phenom_lsr0}).  Such an analysis, however, is predicated on having some knowledge of the $\tau$ 
range for which comparison of the theoretical content of  $R_0(\tau,s_0)$  to (\ref{phenom_lsr0}) is a valid exercise
\cite{SVZ,sr_window}. 
This question can be addressed by H\"older inequalities, which must be upheld for Laplace
sum-rules to retain consistency  with the physical positivity of the resonance content
(\ref{res+cont}) of 
${\rm Im}\Pi^{res}(s)$ \cite{holder}:  
\begin{equation}
\frac{{ R}_0[\omega\tau+(1-
\omega)\delta\tau,s_0]}{\left({R}_0[\tau,s_0]\right)^\omega
\left({ R}_0[\tau+\delta\tau,s_0]\right)^{1-\omega}} \le 1
\quad ,\quad  \forall ~0\le \omega \le 1 \quad .
\label{rat2a}
\end{equation}
If $\delta\tau$ is reasonably small ($\sim 0.1\,{\rm GeV}^{-2}$ \cite{holder}), these inequalities are themselves 
insensitive to the choice of $\delta\tau$, in which case the requirement (\ref{rat2a}) can be used to determine a
range for the Borel parameter $\tau$ itself, given a chosen value for the continuum threshold $s_0$.  Such
H\"older inequalities have in the past been used to place bounds on the light-quark masses and on the 
pion polarizability \cite{phenom_bounds}, as well as the 
parameter space of QCD sum-rule analyses.

\begin{figure}[hbt]
\centering
\includegraphics[scale=0.7,angle=270]{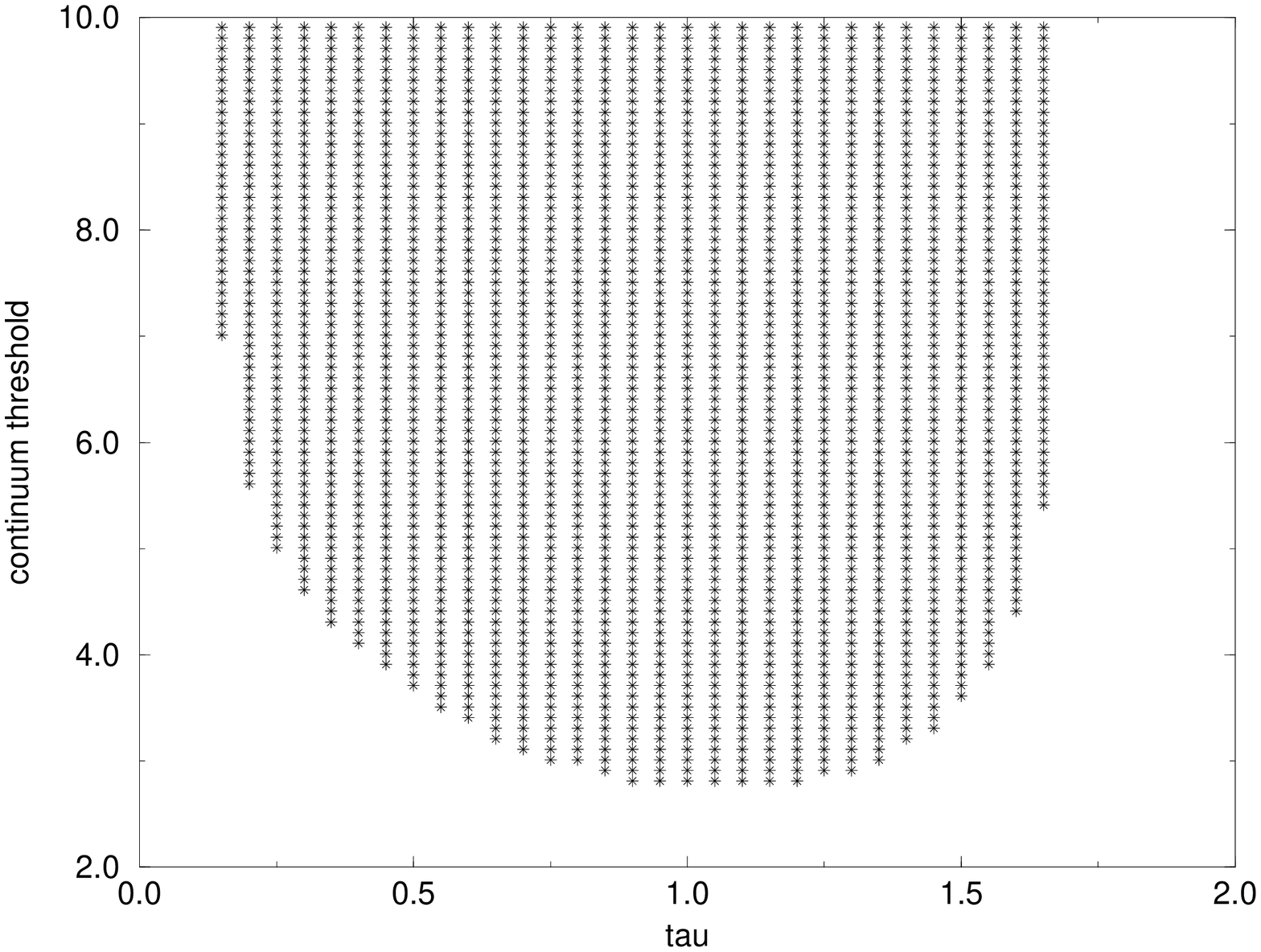}
\caption{Data points indicate values of $\tau$ and $s_0$ for which the 
$I=0$ sum-rule with inclusion of the instanton continuum
satisfies the H\"older inequality.  The scales are in GeV units.
}
\label{ineq_f0_ins}
\end{figure}

\begin{figure}[hbt]
\centering
\includegraphics[scale=0.7,angle=270]{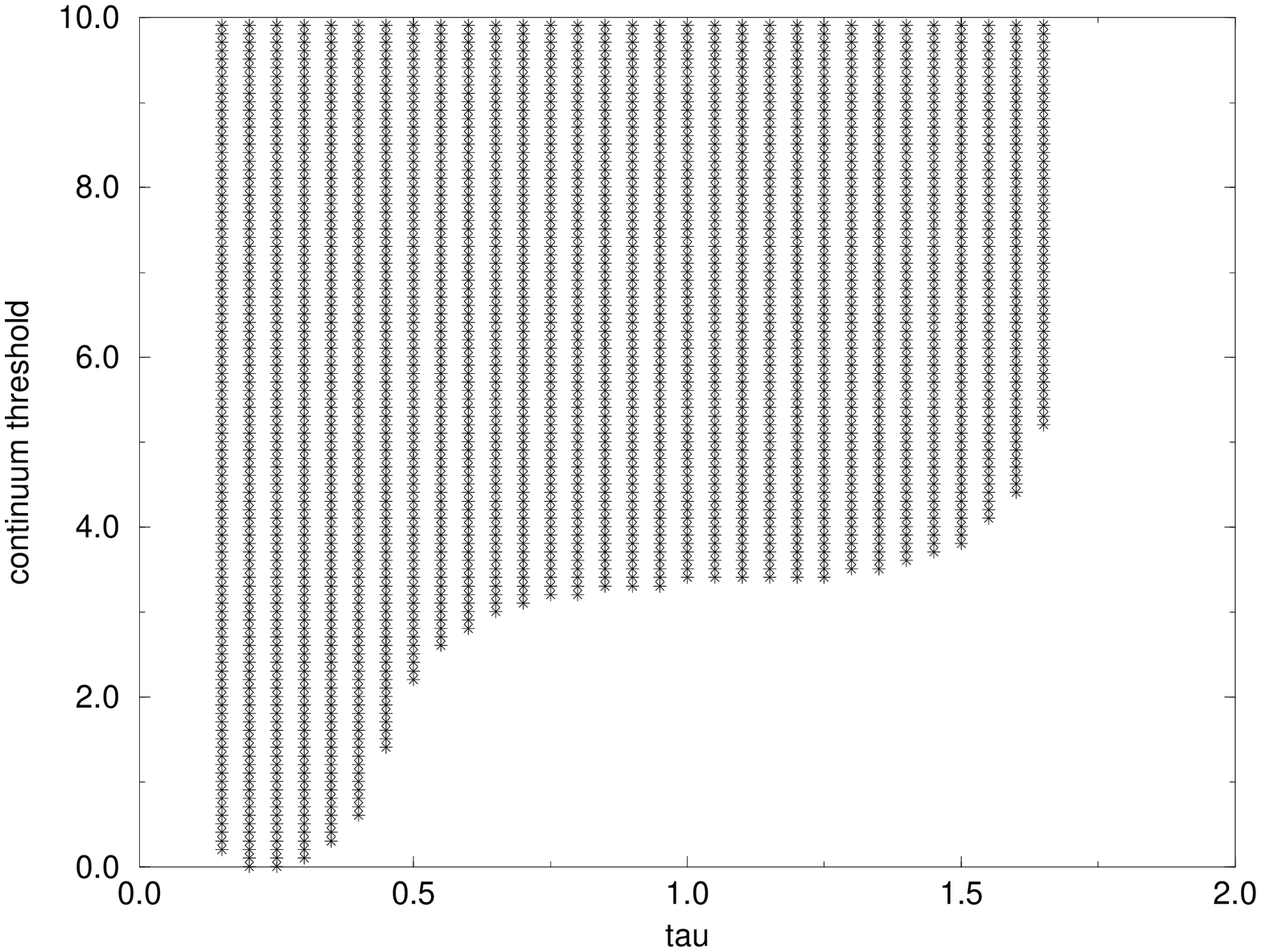}
\caption{
Data points indicate values of $\tau$ and $s_0$ for which the 
$I=0$ sum-rule with omission of the instanton continuum
satisfies the H\"older inequality.  The scales are in GeV units.
}
\label{ineq_f0_noins}
\end{figure}

The inequality (\ref{rat2a}) can be demonstrated to support the presence of an instanton contribution to the continuum 
$c_0(\tau,s_0)$, as this contribution alters the region in $(\tau,s_0)$ parameter space for which (\ref{rat2a}) is
upheld.  Figure \ref{ineq_f0_ins} shows the region of this parameter space satisfying the inequality (\ref{rat2a}) 
when instantons contribute to $c_0(\tau,s_0)$ in the isoscalar case.  The shape of the region is quite similar to that which characterizes other sum rules \cite{holder}.  Omission of this instanton contribution from the 
continuum  leads to the less restrictive parameter space shown in  Figure \ref{ineq_f0_noins}, which seems to imply the 
validity of local duality at very low values of $s_0$ that are generally H\"older-inequality excluded.

\begin{figure}[hbt]
\centering
\includegraphics[scale=0.7,angle=270]{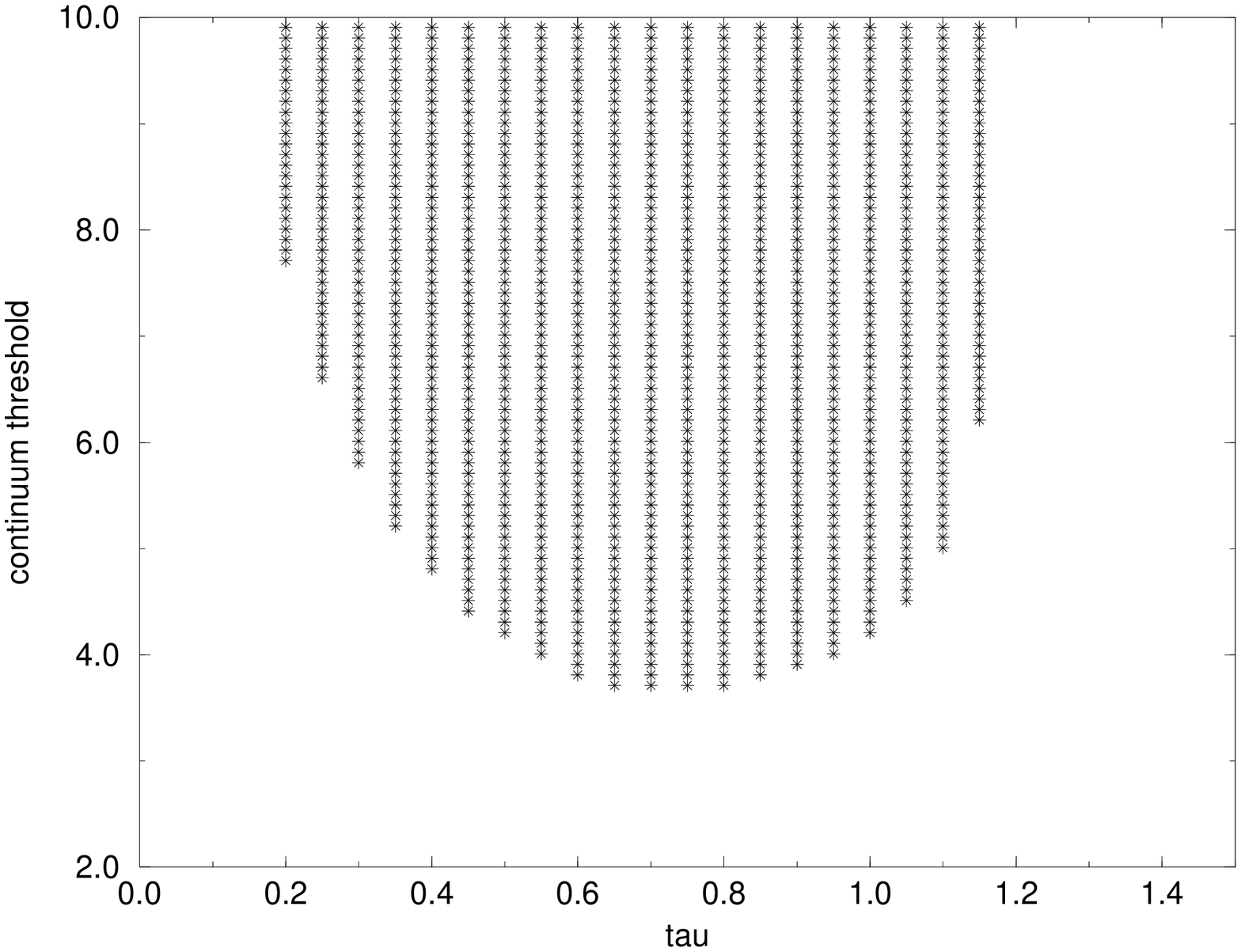}
\caption{
Data points indicate values of $\tau$ and $s_0$ for which the 
$I=1$ sum-rule with inclusion of the instanton continuum
satisfies the H\"older inequality.  The scales are in GeV units.
}
\label{ineq_a0_ins}
\end{figure}

\begin{figure}[hbt]
\centering
\includegraphics[scale=0.7,angle=270]{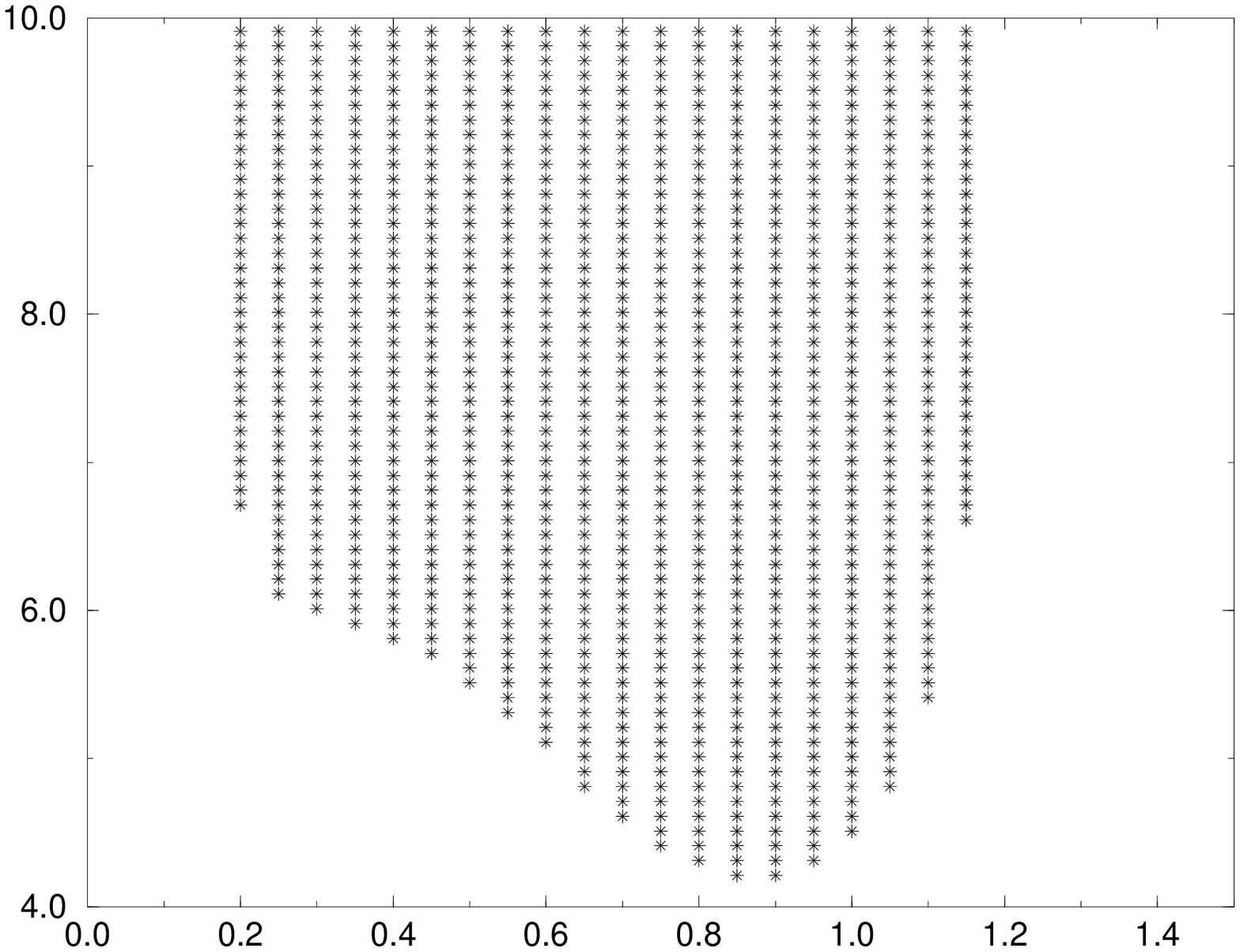}
\caption{
Data points indicate values of $\tau$ and $s_0$ for which the 
$I=1$ sum-rule with omission of the instanton continuum
satisfies the H\"older inequality.  The scales are in GeV units.
}
\label{ineq_a0_noins}
\end{figure}

 Figures \ref{ineq_a0_ins} and \ref{ineq_a0_noins} show the effect of the instanton continuum on
the inequality analysis for the isovector channel.  If the instanton contribution is removed 
(Figure \ref{ineq_a0_noins}), the parameter space for which (\ref{rat2a}) is upheld is restricted 
to substantially larger values of $s_0$ than is the case when the instanton contribution to the continuum 
is retained (Figure \ref{ineq_a0_ins}).
Inclusion of the instanton continuum again leads to behaviour characteristic of other sum-rules \cite{holder}.
Thus we see that if the instanton contribution to the continuum is removed, the $(\tau,s_0)$ parameter space 
allowed by (\ref{rat2a}) increases in the isoscalar channel and decreases in the isovector channel, thereby enhancing
unrealistically the discrepancy between the minimum value of the continuum threshold characterizing these two channels.

Thus the parameter space consistent with (\ref{rat2a}) appears to favour instanton contributions to the continuum,
consistent with the methodology of the previous section.  The allowed regions of the parameter space for the isoscalar
(Figure \ref{ineq_f0_ins}) and isovector (Figure \ref{ineq_a0_ins}) channels are conservatively given by
\begin{eqnarray}
0.3\,{\rm GeV}^{-2}\le \tau\le 1.7\,{\rm GeV}^{-2}~,~s_0>3\,{\rm GeV}^2\quad (I=0),
\label{f0_ineq}
\\
0.3\,{\rm GeV}^{-2}\le \tau\le 1.1\,{\rm GeV}^{-2}~,~s_0>4\,{\rm GeV}^2\quad (I=1).
\label{a0_ineq}
\end{eqnarray}
It should be noted here that the absolute lower bound on $s_0$ in (\ref{f0_ineq})
excludes the $s_0\le1.6\,{\rm GeV^2}$ range, as determined in Section \ref{fesr_theor_sec}, for which FESR's allow
a light $n\bar n$ $\sigma$ resonance to be masked by the heavier $f_0(980)$.  However, the possibility that such FESR
masking of a light $n \bar n$ $\sigma$ occurs via the heavier $f_0(1370)$, as discussed at the end of Section \ref{fesr_theor_sec}, is still
marginally allowed by the H\"older inequality (\ref{rat2a}).  Recall from Section \ref{fesr_theor_sec} that for such a masking to occur, the
necessary condition that $\left(F_1/F_0\right)^{1/2}\le 1.37\,{\rm GeV}$ is upheld provided $s_0< 3.5\,{\rm GeV^2}$, an
upper bound that is seen to be near the very bottom of the permitted parameter space in Figure \ref{ineq_f0_ins}.  Consequently, 
such a scenario cannot at this juncture be ruled out, and requires exploration by Laplace sum rules
over H\"older-inequality allowed parameter space.  

Finally, we note that the upper bound on the $\tau$-range in (\ref{f0_ineq}) insures that
exponential factors in (\ref{phenom_lsr0}) will not suppress any isoscalar resonance lighter than $800\,{\rm MeV}$, ensuring
that the Laplace sum-rule remains an appropriate tool for extracting a 400--700 MeV $\sigma$ resonance.  Moreover, the upper bound in 
(\ref{a0_ineq}) similarly ensures that an isovector resonance lighter than $1\,{\rm GeV}$ will also not be subject to 
exponential suppression, in which case the sum-rule  in the isovector channel should be sensitive to $a_0(980)$ contributions.
We shall see in the section that follows that {\em neither} of these candidates for the lowest-lying $n\bar n$ resonance in 
their respective channels are evident from the Laplace sum-rule $R_0(\tau,s_0)$, suggesting that both the $\sigma(500)$ 
[if such a state exists at all] and the $a_0(980)$ should be interpreted either as {\em non} $q\bar q$ resonance states
or as $q\bar q$ states with negligible $n\bar n$ content.

\section{Phenomenological Analysis of Laplace Sum-Rules}
\label{phenom_lap_sec}
\subsection{Single-Resonance Analysis}
\label{one_res_sec}
In Section \ref{laplace_theor_sec} it was also shown that we can anticipate enhanced splitting between isopartner states resulting 
from an instanton contribution to the continuum, suggesting a need to revisit the sum-rule analysis of 
\cite{first_paper} for the $n\bar n$ component of the lowest-lying $I=\{0,1\}$ quark ($q\bar q$) scalar mesons.
The sum-rule predictions of the properties of these lowest-lying $I=\{0,1\}$ $n\bar n$ scalar resonances can now
be studied through  (\ref{res_sr}), which is assumed for now to be dominated by the lowest-lying resonance 
(as discussed at the end of Section \ref{fesr_theor_sec}).  
Since the widths of such resonances may be substantial, it is necessary to extend the narrow width approximation 
traditionally used in sum-rules.   A flexible and numerically simple technique  is 
to build up the resonance shape (\ref{res_dist}) within $Im\Pi(s)$ [eq. (\ref{res+cont})]  by utilizing $n$
 unit-area square 
pulses  \cite{first_paper,width}, thereby replacing (\ref{res_dist}) with the following expression:
\begin{eqnarray}
S_r^{(n)}(s) &=&  \frac{2}{n\pi} \sum_{j=1}^n \sqrt{ \frac{n-j+f}{j-f}} P_{M_r} \left[
s, \sqrt{\frac{n-j+f}{j-f}} \; \Gamma_r \right]\quad ,
\label{n-pulse}
\\
P_M\left[s,\Gamma\right]&\equiv&\frac{1}{2M\Gamma}\left[\Theta(s-M^2+M\Gamma)-\Theta(s-M^2-M\Gamma)\right]
\quad .
\label{square_pulse}
\end{eqnarray}
A single square pulse models a broad nearly structureless contribution (such as a broad light $\sigma$) to ${\rm Im}\Pi(s)$,
while a Breit-Wigner resonance of a particle of mass $M$ and width $\Gamma$  can be expressed as a sum of a number of square pulses.
The quantity $f$ can be fixed by normalizing the area of the n-pulse approximation to unity.

We begin the phenomenological analysis with the  $n=4$ pulse approximation (\ref{n-pulse})
to the resonance shape  so that the right-hand side of (\ref{res_sr}) becomes
\begin{eqnarray}
& &\frac{1}{\pi}{\rm Im}\Pi^{res}=g_rS_r^{(4)}(s)\quad ,
\label{res_shape_def}\\
& &{R}_0(\tau,s_0)=g_re^{-M^2\tau}W_4(M,\Gamma,\tau)\quad ,\label{basic_fit}\\
& &W_4(M,\Gamma,\tau)=
\frac{2}{4\pi} \sum_{j=1}^4 \frac{1}{M\Gamma\tau}\sinh\left[M\sqrt{\frac{4-j+0.7}{j-0.7}}\,\Gamma\tau\right]
\quad ,
\label{4-pulse-W}
\end{eqnarray}
where $\sqrt{g_r}$ is proportional to the strength with which the scalar $n\bar n$ current couples the vacuum to the resonance.
As noted earlier, we are ignoring all but the lowest-lying resonance contribution to (\ref{res+cont}) in part
because of the anticipated exponential suppression of subsequent resonances [{\it e.g.} (\ref{phenom_lsr0}) ].  
The free parameters in the relation (\ref{basic_fit}), the resonance-related quantities $g_r$, $M$, $\Gamma$ and the continuum-threshold 
$s_0$, can be extracted from a fit to the $\tau$ dependence of the theoretical expression
${R}_0(\tau,s_0)$.  This is done by minimizing the $\chi^2$ defined by
\begin{equation}
\chi^2=\frac{1}{N}\sum_{j=1}^N\frac{\left[{R}_0\left(\tau_j,s_0\right)-g_re^{-M^2\tau_j}W_4(M,\Gamma,\tau_j)\right]^2}{\epsilon(\tau_j)^2}
\quad ,
\label{chi2}
\end{equation}  
where the sum is over evenly spaced, discrete $\tau$ points in the ranges (\ref{f0_ineq},\ref{a0_ineq}) consistent with the
H\"older inequality.  The weighting factor $\epsilon$ used for the evaluation of (\ref{chi2}) is
$\epsilon(\tau)=0.2R_0(\tau,s_0)$.
This 20\% uncertainty has the desired property of being dominated by the continuum at low $\tau$ and power-law corrections at large
$\tau$.  Other choices of the $0.2$ prefactor in $\epsilon$ would simply rescale the $\chi^2$, so its choice has no effect on
the values of the $\chi^2$-minimizing parameters.

In the $\chi^2$ minimization, the RG-invariant quark mass parameter $\hat m_q$ (\ref{hatm}) is now absorbed into the quantity
$a=g_r/\hat m_q^2$. The best-fit parameters for the $I=\{0,1\}$ channels are  shown in Table \ref{best_tab}. Since the resonance widths are sufficiently small to permit
a series expansion of the $n$-pulse approximation ({\it i.e.} $M\Gamma\tau$ is small),
it is impossible for $\chi^2$-minimization to distinguish between the $n=4$ and $n=1$ approximations;  
each has the same $\tau$ dependence to second order in their series 
expansions.  Thus there is no possibility of distinguishing between a structureless resonance shape 
represented by $n=1$ and a Breit-Wigner-like form represented by $n=4$. We will continue
to use the four-pulse approximation since it will lead to values for $\Gamma$ that 
correspond to a Breit-Wigner width \cite{width}.

\begin{table}[hbt]
\begin{tabular}{||c|c|c|c|c||}\hline\hline
$I$ & $M~(GeV)$ & $s_0~(GeV^2)$ & $a~(GeV^4)$ & $\Gamma ~(GeV)$
\\ \hline\hline
0  &  $0.97$ & $3.6$ & $0.073$ & $0.24$ 
\\ \hline
1 & $1.50$  & $4.5$  &$0.16$ & $0.20$
\\\hline\hline
\end{tabular}
\caption{  
 Values
 for the resonance parameters and continuum threshold for the $I=\{0,1\}$ channels
which lead to a minimum $\chi^2$.
}
\label{best_tab}
\end{table}

In principle the $2\pi$ continuum should also be included in the phenomenological model.  
However, the values for $a$ in Table \ref{best_tab} lead to a resonance contribution 
which is much larger than the $2\pi$ continuum \cite{LRT}: 
\begin{equation}
\frac{1}{\hat m_q^2}\frac{1}{\pi}{\rm Im}\,\Pi^{2\pi}(t)=\frac{3}{16\pi^2}\left\vert
F(t)\right\vert^2\theta\left(t-4m_\pi^2\right)
\quad ,\quad F(t)\approx m_\pi^2\left(1+\frac{1}{6}\langle r_\pi^2\rangle t\right)
\quad .
\end{equation}

To determine the uncertainties associated with the best-fit parameters 
of Table \ref{best_tab} we perform a 
 Monte-Carlo simulation which includes  the 
parameter ranges $1\le f_{vs}\le 2$ and a 15\% variation in the instanton size $\rho$.
Using the technique of \cite{Steele}, we also  simulate  continuum and OPE-truncation
uncertainties from the empirical functions 
\begin{equation}
\epsilon(\tau)=a_{1}\exp{\left[-1.1\tau^{0.73}\right]} \quad , \quad I=0 \quad ;\quad
\epsilon (\tau)=a_{2}\tau \exp{\left[-5.7\tau^{0.53}\right]} \quad ,\quad I=1\quad ,
\label{err_mod}
\end{equation}
where $\vert a_1\vert \le 0.17$ and $\vert a_2\vert \le 1.15$ in the Monte-Carlo simulation.

The Monte-Carlo simulation leads to  the 90\% confidence level results for the best-fit parameters 
shown in   Table \ref{mc_tab}.  
These results indicate an absence of both the isovector $a_0(980)$ resonance as well as a very broad
$400$--$700\,{\rm MeV}$
isoscalar $\sigma$ state, suggesting that neither of these states are primarily  
$n\bar n$ mesons; both appear to be
decoupled from sum-rules based upon the $n\bar n$ scalar current (\ref{current}). 
As mentioned earlier, decreasing the number of pulses (to simulate a structureless resonance) does not alter the $\chi^2$,
 and only leads to a rescaling of $\Gamma$.  
The large uncertainty in the width $\Gamma$ 
indicates the difficulty associated with predicting widths from QCD sum-rules, a problem which merits future consideration. 
An alternative choice for the resonance model will be considered in 
Section \ref{shape_sec}.

\begin{table}[hbt]
\begin{tabular}{||c|c|c|c|c||}\hline\hline
$I$ & $M~(GeV)$ & $s_0~(GeV^2)$ & $a~(GeV^4)$ & $\Gamma ~(GeV)$
\\ \hline\hline
0  &  $1.00\pm 0.09$ & $3.7\pm 0.4$ & $0.08\pm 0.02$ & $0.19\pm 0.14$ 
\\ \hline
1 & $1.55\pm 0.11$  & $5.0\pm 0.7$  &$0.17\pm 0.04$ & $0.22\pm 0.11$
\\\hline\hline
\end{tabular}
\caption{{  Results of the Monte-Carlo simulation of 90\% confidence-level uncertainties 
 for the resonance parameters and continuum threshold for the $I=\{0,1\}$ channels.} 
}
\label{mc_tab}
\end{table}

\subsection{Multi-Resonance Scenarios}
\label{multi_res_sec}
A simple extension of the single resonance model (\ref{basic_fit}) is a model with the 
incoherent sum
of two resonances.  The best fits in such a model lead  to results which either 
are degenerate to
the single resonance values of Table \ref{best_tab}, or which do not significantly 
reduce the $\chi^2$ compared with the single resonance model.  The latter
case indicates that the second resonance is weak enough to be absorbed into
the continuum, implying a decoupling from the $n\bar n$ scalar current.

To  study quantitatively  this apparent decoupling of a light $\sigma$ and
the $a_0(980)$ from the quark ($n\bar n$) scalar current, we employ a two-resonance version
of (\ref{basic_fit}) in which the masses and widths of the resonances are used
as input parameters. The two-resonance model couplings ({\it i.e.} the parameter $a=g_r/\hat
m_q^2$ for each resonance) and continuum threshold $s_0$ which minimize  $\chi^2$ are
explicitly calculated.  

As input we utilize the PDG values \cite{PDG} for the $f_0(980)$, 
$a_0(980)$, $a_0(1450)$, as well as a $\sigma$ with $M=500\,{\rm MeV}$ and
$\Gamma=500\,{\rm MeV}$.  Since the $f_0(980)$ and $a_0(1450)$ PDG values are
consistent 
with the fits of Table \ref{mc_tab}, we anticipate that  if  $a_0(980)$ and $\sigma$ are
truly decoupled from the $n\bar n$ quark scalar currents,  then the $\chi^2$-minimizing
values of the $a_0(1450)$
coupling,
$f_0(980)$ coupling, and $s_0$ should reproduce those given in Table \ref{best_tab}.
Although the observed hadronic states could in general contain a mixture of $n\bar n$, $s\bar s$, 
 and possibly  meson-meson
or four-quark   
components, any state with a non-zero $n\bar n$ component will couple optimally  to
the scalar current (\ref{current}) [the more exotic components may couple to 
(\ref{current}) as well, but to a lesser degree].  
Consequently, a large suppression of a resonance within the context of an $n\bar n$-current
sum-rule analysis is indicative of a reduction of that same resonance's explicit $n\bar n$ content.

Table \ref{multi_tab} displays the results of this analysis.  The fitted
values of the $a_0(1450)$  coupling, the $f_0(980)$ coupling, and $s_0$ are
virtually identical to those of Table \ref{best_tab}.  
Furthermore, Table \ref{multi_tab}
shows that the $\sigma$ and $a_0(980)$ have couplings to the $n\bar n$ scalar current 
which are significantly suppressed compared with those of the $f_0(980)$ and 
$a_0(1450)$.  

Since the FESR results of Section \ref{fesr_theor_sec} suggest the possibility of a light $\sigma$ hiding beneath
the $f_0(1370)$, we extend the isoscalar results of Table \ref{multi_tab} by including the 
$f_0(1370)$.  The results of this extension  shown in Tables \ref{multi_tab2} and \ref{multi_tab3}
indicate a suppression of the $\sigma$ and $f_0(1370)$ couplings to the $n\bar n$ current 
(\ref{current}) compared with that of
the $f_0(980)$.

\begin{table}[hbt]
\begin{tabular}{||c|c|c|c|c|c||}\hline\hline
$I$ & $s_0~(GeV^2)$ & State & $M~(GeV)$ & $\Gamma ~(GeV)$ & $a~(GeV^4)$  
\\ \hline\hline
1 & $4.4$ &  $a_0(980)$ & $0.98$ & $0.1$ & $0.0075$ 
\\ \hline
1 & $4.4$ &  $a_0(1450)$ & $1.45$ & $0.2$ & $0.16$ 
\\ \hline
0 & $3.7$ &  $\sigma$ & $0.5$ & $0.5$ & $0.0051$ 
\\ \hline 
0 & $3.7$ &  $f_0(980)$ & $0.98$ & $0.1$ & $0.068$ 
\\\hline\hline
\end{tabular}
\caption{{  
Best fit values for the couplings and $s_0$ in a multi-resonance scenario.
The masses and widths were used as input values.}}
\label{multi_tab}
\end{table}

\begin{table}[hbt]
\begin{tabular}{||c|c|c|c|c||}\hline\hline
 $s_0~(GeV^2)$ & State & $M~(GeV)$ & $\Gamma ~(GeV)$ & $a~(GeV^4)$  
\\ \hline\hline
 $3.8$ &  $\sigma$ & $0.5$ & $0.5$ & $0.0057$ 
\\ \hline 
 $3.8$ &  $f_0(980)$ & $0.98$ & $0.1$ & $0.065$ 
\\ \hline
$3.8$ & $f_0(1370)$ & $1.37$ & $0.2$ & $0.0055$
\\\hline\hline
\end{tabular}
\caption{{  
Best fit values for the couplings and $s_0$ in a multi-resonance scenario for the
isoscalar channel with the PDG \protect\cite{PDG} lower bound on the $f_0(1370)$ width.
} 
}
\label{multi_tab2}
\end{table}

\begin{table}[hbt]
\begin{tabular}{||c|c|c|c|c||}\hline\hline
 $s_0~(GeV^2)$ & State & $M~(GeV)$ & $\Gamma ~(GeV)$ & $a~(GeV^4)$  
\\ \hline\hline
 $3.7$ &  $\sigma$ & $0.5$ & $0.5$ & $0.0051$ 
\\ \hline 
 $3.7$ &  $f_0(980)$ & $0.98$ & $0.1$ & $0.068$ 
\\ \hline
$3.7$ & $f_0(1370)$ & $1.37$ & $0.5$ & $0.00098$
\\\hline\hline
\end{tabular}
\caption{{  
Best fit values for the couplings and $s_0$ in a multi-resonance scenario for the
isoscalar channel with the PDG \protect\cite{PDG} upper bound on the $f_0(1370)$ width.
} 
}
\label{multi_tab3}
\end{table}

 The above results can be interpreted by recalling that the parameter $a$ is related to the strength of the 
coupling of the hadronic state $\vert H\rangle$ to the vacuum via the 
$n\bar n$ current.
\begin{equation}
a\sim \left\vert \langle O\vert n\bar n\vert H\rangle\right\vert^2
\label{coulping_def}
\end{equation}
Hence it is clear from Tables \ref{multi_tab}--\ref{multi_tab3} that the $\sigma$ and $a_0(980)$
have substantially reduced coupling to the $n\bar n$ current than the dominant states $f_0(980)$
and $a_0(1450)$, suggesting a non-$n\bar n$ interpretation of the $\sigma$ and $a_0(980)$. For example,
the coupling of a meson-meson state to the vacuum through a $q\bar q$ current is likely to be suppressed
compared to that of a $q\bar q$ state.  Certainly at the QCD level, the mixing of local 
$q\bar q$ and $qq\bar q\bar q$ currents will be chirally suppressed by quark mass factors, which
suggests a theoretical basis for this decoupling.

The relative strength of the couplings obtained in Tables  \ref{multi_tab}--\ref{multi_tab3}
can be used as an estimate of the $n\bar n$ content of the known hadronic states.  
If the dominant states 
 with the largest coupling are nearly pure $n\bar n$, and  if there is minimal 
interference in the mixing between the $n\bar n$ and other components 
(such as meson-meson or four-quark components)
of the non-dominant states, then   Tables \ref{multi_tab}--\ref{multi_tab3} 
indicate that the $a_0(980)$ 
has a relative  $n\bar n$-content given approximately by
$\sqrt{0.0075/0.16}\approx 20\%$, and that the $n\bar n$ content of the $\sigma$ 
is approximately  25--30\%.  If the dominant states are themselves admixtures, then the 
$n\bar n$-content of the $a_0(980)$ and $\sigma$ is correspondingly reduced.

Another important consequence of this analysis is the clear distinction 
that emerges between  $f_0(980)$ and $a_0(980)$, since the former 
appears as the dominant state,  and the latter appears to be  coupled only weakly in 
comparison with the dominant isovector-channel state $a_0(1450)$.    Such
a distinction is important, since both $980\,{\rm MeV}$ states are close to the $K\bar K$ 
kinematic threshold.  Consequently,  both states might be expected
to be dominated by comparably large  $K\bar K$ components, in which case both states
would exhibit comparably suppressed couplings to $n\bar n$-current scalar-channel sum rules.
The disparity we find, however, is indicative of substantial $n\bar n$ content in $f_0(980)$,
a result with some experimental support \cite{Lafferty}.

\subsection{Higher-Loop Perturbative Effects}
\label{hl_sec}
The effect of higher order perturbative contributions  
on the resonance parameters can now be studied.  The 
perturbative part of the
correlation function is known to four-loop order \cite{Chetyrkin}, modifying  (\ref{laplace_theor}) as follows: 
\begin{eqnarray}
{ R}_0(\tau,\infty)&=&\frac{3m_q^2}{16\pi^2\tau^2}\left(   
1+4.821098 \frac{\alpha_s}{\pi}+21.97646\left(\frac{\alpha_s}{\pi}\right)^2+53.14179\left(\frac{\alpha_s}{\pi}\right)^3
\right)
\label{srhl}
\\
& +&m_q^2\left(
\frac{3}{2}\langle m_q\bar q q\rangle 
+\frac{1}{16\pi}\langle \alpha_s G^2\rangle
+\pi\langle{\cal O}_6\rangle \tau
\right)
+\left(-1\right)^Im_q^2
{3\rho^2\over{16 \pi^2\tau^3}} e^{-\frac{\rho^2}{2\tau} }
\left[   
  K_0\left( \frac{\rho^2}{2\tau} \right) +
       K_1\left(\frac{\rho^2}{2\tau}  \right)
\right].
\nonumber
\end{eqnarray}
Similarly, the continuum contributions (\ref{pert_cont}) to four-loop order become
\begin{eqnarray}
& &c_{0}^{pert}\left(\tau,s_0\right)=\frac{1}{\pi}\int\limits_{s_0}^\infty Im\Pi^{pert}(s) e^{-s\tau}\,ds
\quad ,
\\
& &c_0^{{pert}}\left(\tau,s_0\right)
= \frac{3m_q^2}{16\pi^2}\left[A_0f_0\left(\tau, s_0\right)+A_1f_1\left(\tau, s_0\right)+A_2f_2\left(\tau, s_0\right)
+A_3f_3\left(\tau, s_0\right)\right]
\label{pert_cont_hl}\quad ,\\
& &A_0=1+\frac{17}{3}\frac{\alpha_s}{\pi}+31.8640\left(\frac{\alpha_s}{\pi}\right)^2+89.1564\left(\frac{\alpha_s}{\pi}\right)^3
~ ,~ A_1=-2\frac{\alpha_s}{\pi}-\frac{95}{3}\left(\frac{\alpha_s}{\pi}\right)^2-297.596\left(\frac{\alpha_s}{\pi}\right)^3
~,
 \\ 
& &A_2=\frac{17}{4}\left(\frac{\alpha_s}{\pi}\right)^2+\frac{229}{2}\left(\frac{\alpha_s}{\pi}\right)^3
\quad ,\quad
A_3=-\frac{221}{24}\left(\frac{\alpha_s}{\pi}\right)^3\quad ,
\\
& &f_2\left(\tau,s_0\right)=\frac{1}{\tau^2}\biggl[2\log\left(s_0\tau\right)\left(e^{-s_0\tau}+E_1\left(s_0\tau\right)-1+\gamma_E\right)
+\log^2\left(s_0\tau\right)\left[1+\left(1+s_0\tau\right)e^{-s_0\tau}\right]\biggr.
\nonumber\\
& &\qquad\qquad\qquad\qquad
\biggl.-\frac{1}{4}\left(s_0\tau\right)^2
\phantom{\,}_3F_3\left(2,2,2;3,3,3;-s_0\tau \right)+\frac{\pi^2}{6}-2\gamma_E+\gamma^2\biggr]
\quad ,
\label{f2}
\\
& &f_3\left(\tau,s_0\right)=\frac{1}{\tau^2}\Biggl[
\log^3\left(s_0\tau\right)\left[2+\left(1+s_0\tau\right)e^{-s_0\tau}\right]+
3\log^2\left(s_0\tau\right)\left[e^{-s_0\tau}+E_1\left(s_0\tau\right)-1+\gamma_E\right]
\Biggr.
\nonumber\\
& &\qquad\qquad\qquad\qquad
-2\zeta(3)+\frac{\pi^2}{2}-\frac{1}{2}\pi^2\gamma_E+3\gamma_E^2-\gamma_E^3
+\frac{3}{8}\left(s_0\tau\right)^2\phantom{\,}_4F_4\left(2,2,2,2;3,3,3,3;-s_0\tau \right)
\nonumber\\
& &\qquad\qquad\qquad\qquad\Biggl.
-\frac{3}{4}\left(s_0\tau\right)^2\log\left(s_0\tau\right)
\phantom{\,}_3F_3\left(2,2,2;3,3,3;-s_0\tau \right)
\Biggr]\quad ,
\label{f3}
\end{eqnarray}
where 
$\phantom{\,}_pF_q$ 
represents the generalized Hypergeometric function \cite{abram}. 
Finally, it is necessary to utilize the running coupling constant and running mass to 
four-loop order in the perturbative corrections.
The four-loop ($n_f=3$) result for the running coupling constant is \cite{beta,run_alpha}
\begin{eqnarray}
\frac{\alpha_s(\mu)}{\pi}&=& \frac{1}{\beta_0 L}-\frac{\bar\beta_1\log L}{\left(\beta_0L\right)^2}+
\frac{1}{\left(\beta_0 L\right)^3}\left[
\bar\beta_1^2\left(\log^2 L-\log L -1\right) +\bar\beta_2\right]
\nonumber\\
&+ &\frac{1}{\left(\beta_0 L\right)^4}\left[\bar\beta_1^3\left(-\log^3 L+\frac{5}{2}\log^2 L+2\log L -\frac{1}{2}\right)
-3\bar\beta_1\bar\beta_2\log L+\frac{\bar\beta_3}{2}\right]
\quad ,
\label{alpha_hl}\\
& &L=\log\left(\frac{\mu^2}{\Lambda^2}\right)\quad ,\quad \bar\beta_i=\frac{\beta_i}{\beta_0}
\quad ,
\nonumber\\
& &\beta_0=\frac{9}{4}\quad ,\quad \beta_1=4\quad ,\quad \beta_2=\frac{3863}{384}\quad ,\quad 
\beta_3=\frac{445}{32}\zeta(3)+\frac{140599}{4608}\quad ,
\end{eqnarray}
where $\zeta(n)$ is the Zeta function. 
Similarly, the  four-loop ($n_f=3$) result for the running quark mass is \cite{run_mass}
\begin{eqnarray}
m(\mu)&\equiv& \hat m G(\mu) \quad ,
\\
G(\mu)&=&\frac{1}{\left(\frac{1}{2}L\right)^{\frac{4}{9}}}\biggl[
1+\frac{290}{729}\frac{1}{L}-\frac{256}{729}\frac{\log{ L}}{L}
\biggr.
+\left(\frac{550435}{1062882}-\frac{80}{729}\zeta(3)\right)\frac{1}{L^2}
 -\frac{388736}{531441}\frac{\log{L}}{L^2}+\frac{106496}{531441}\frac{\log^2{L}}{L^2}
\nonumber\\
& &\qquad\qquad
\left(\frac{2121723161}{2324522934}+\frac{8}{6561}\pi^4-\frac{119840}{531441}\zeta(3)-\frac{8000}{59049}\zeta(5)\right)\frac{1}{L^3}
\label{mass_hl}\\
& &\qquad\qquad
\biggl.
\left(-\frac{611418176}{387420489}+\frac{112640}{531441}\zeta(3)\right)\frac{\log{L}}{L^3}
+\frac{335011840}{387420489}\frac{\log^2{L}}{L^3}-\frac{149946368}{1162261467}\frac{\log^3{L}}{L^3}\biggr]\quad .
\nonumber
\end{eqnarray}

The higher-loop perturbative effects are accompanied by numerically large coefficients, which raise concerns about the convergence of the perturbation series for the scalar correlator.  However, it seems reasonable that the nonperturbative corrections should 
similarly be accompanied by large higher-loop  corrections, so the inclusion of  
only higher-loop perturbative effects may artificially 
enhance the size of purely perturbative contributions relative to nonperturbative 
(condensate and instanton) contributions.
With this reservation, we proceed  to study the stability of the sum-rule predictions
against such higher-order perturbative corrections.  

Table \ref{higher_tab} shows the effect of higher-loop
perturbative corrections on the $\chi^2$-minimizing resonance parameters.  The minimum $\chi^2$ does not
change significantly as higher-loop perturbative effects are included.
The mass $M$ and coupling $a$ are remarkably stable under higher-loop corrections, 
a consistency already apparent from comparison of 2-loop and 4-loop 
perturbative contributions to FESR ratios related to the resonance mass (Figure \ref{fesr_ratio_fig}), 
while the continuum threshold $s_0$ decreases and the resonance width increases with increasing order of perturbation theory.
Such results are not difficult to understand. The decrease in $s_0$  cancels a portion of the perturbative 
contribution to $R_0$, as evident from (\ref{laplace_theor}).  
Furthermore, since both the width factor $W_4$  and nonperturbative effects increase with increasing $\tau$
(in contrast to perturbative effects which decrease with increasing $\tau$),
 the increase in the width compensates for the  artificially diminished role of nonperturbative contributions as
the perturbative contributions are taken to higher-loop orders.

It seems reasonable to expect that with moderate (unknown) higher-order nonperturbative corrections, a Monte-Carlo 
simulation of uncertainties  would lead to phenomenological predictions similar to Table \ref{mc_tab}, except for a  
decrease in the central value of the continuum threshold $s_0$, as is observed in other sum-rule analyses \cite{width}.  
Since the width $\Gamma$ is seen to be highly uncertain in 
Table \ref{mc_tab}, it is not clear 
whether  higher-loop perturbative and nonperturbative effects would lead to a 
significant change in the central value  of $\Gamma$.

\begin{table}[hbt]
\begin{tabular}{||c|c|c|c|c|c||}\hline\hline
$I$& Loop Order & $M~(GeV)$ & $s_0~(GeV^2)$ & $a~(GeV^4)$ & $\Gamma ~(GeV)$
\\ \hline\hline
 0 & 2 &  $0.97$ & $3.6$ & $0.073$ & $0.24$ 
\\ \hline
 0 & 3 &  $0.96$ & $2.8$ & $0.086$ & $0.35$ 
\\ \hline
 0 & 4 &  $0.94$ & $2.2$ & $0.083$ & $0.45$ 
\\ \hline
1 &2& $1.50$  & $4.5$  &$0.16$ & $0.20$
\\\hline
1 &3& $1.51$  & $4.1$  &$0.17$ & $0.36$
\\\hline
1 &4& $1.55$  & $3.8$  &$0.19$ & $0.45$
\\\hline\hline
\end{tabular}
\caption{  
 Values
 for the resonance parameters and continuum threshold for the $I=\{0,1\}$ channels
which lead to a minimum $\chi^2$.
 }
\label{higher_tab}
\end{table}
 
\subsection{Width-Independent Mass Bounds on the Lightest Scalar States}
\label{bounds_sec}
If the $\chi^2$-minimization criterion used to predict the resonance
parameters from the $\tau$ dependence of $R_0(\tau,s_0)$ is abandoned, 
then it is still possible to obtain bounds on the masses of the $I=\{0,1\}$
states.  These bounds are interesting since they are independent of the
resonance shape and width in a wide class of models, and hence provide
independent support for the conclusions of the previous sections.
These bounds involve use a higher-moment sum-rule
\begin{equation}
{R}_1\left(\tau,s_0\right)\equiv { R}_1(\tau,\infty)-c_1\left(\tau,s_0\right)
=\frac{1}{\pi}\int\limits_0^{s_0}s\, Im\Pi^{res}(s) e^{-s\tau}\,ds
\label{res_sr1}
\end{equation} 
in conjunction with (\ref{res_sr}).
For a {\em single} narrow resonance
[{\it e.g.} for (\ref{res_shape_def}) with $S_r(s)=\delta(s-M^2)$],
the mass $M$ of the lightest state is just
\begin{equation}
M^2=\frac{R_1\left(\tau,s_0\right)}{{R}_0\left(\tau,s_0\right)}\quad .
\label{ratio}
\end{equation}
This technique was employed in \cite{first_paper} to study the $I=\{0,1\}$ scalar mesons.

To extend this method beyond the narrow resonance approximation we  consider
via (\ref{res_sr1}) the identity
\begin{eqnarray}
R_1(\tau,s_0)&=&
M^2\int\limits_0^{s_0} e^{-s\tau} \frac{1}{\pi}Im\Pi^{res}(s)\,ds
+\int\limits_0^{s_0} (s-M^2)e^{-s\tau} \frac{1}{\pi}Im\Pi^{res}(s)\,ds
\nonumber
\\
&=&M^2 R_0(\tau,s_0)+
\int\limits_0^{s_0}(s-M^2)e^{-s\tau} \frac{1}{\pi}Im\Pi^{res}(s)\,ds
\quad ,
\label{bounds2}
\end{eqnarray} 
where $M^2$ represents the mass of the single resonance where $Im\Pi^{res}(s)$ peaks.
If the second term on the right-hand side of (\ref{bounds2})
is negative, then we see that
\begin{equation}
M^2\ge \frac{{ R}_1\left(\tau,s_0\right)}{{R}_0\left(\tau,s_0\right)}\quad .
\label{ratio_bound}
\end{equation}
The negativity of the second term is a reasonable assumption,  since the
overall sign of this integral is sensitive to  the asymmetry
 of the quantity $e^{-s\tau} Im\Pi^{res}(s)$
about the resonance peak $s=M^2$.  If the resonance shape is symmetric (or only mildly
asymmetric) and if the peak is well-contained below the continuum threshold, 
then the exponential weight suppresses the $s>M^2$ region
compared with the $s<M^2$ region, leading to a negative contribution to the
final integral in (\ref{bounds2}).  

The validity of this constraint, and hence the validity of the bound
(\ref{ratio_bound}),   has been explicitly verified in 
a variety of single resonance models, including a broad 
structureless object represented by a single square pulse, a Breit-Wigner shape, 
$n$-pulse approximations to a Breit-Wigner shape, and a Gaussian.  
In particular, it is found in \cite{first_paper} that the mass explicitly 
increases with width, 
implying that 
the lowest-lying state cannot be simultaneously light and wide.

A lower bound on the masses of the $I=0$ and $I=1$ states can now 
be obtained by
calculating the minimum value of the ratio on the right-hand side of (\ref{ratio_bound}) over
the regions of $(\tau,s_0)$ parameter space consistent with the H\"older
inequality as given by Figures \ref{ineq_f0_ins} and \ref{ineq_a0_ins}. 
This gives the lowest possible mass in a single resonance model which is
consistent with the fundamental constraints imposed by the H\"older
inequality.
The resulting mass bounds are 
\begin{equation}
M_{I=0}\ge 840\,{\rm MeV}\quad ,\quad M_{I=1}\ge 1.36\,{\rm GeV}\quad .
\label{mass_bounds}
\end{equation}     

These mass bounds must be interpreted with care since they are only valid in 
a model containing a single resonance.   
As shown in the multi-resonance scenarios,
it is  possible to have a  resonance which is weakly coupled
to the scalar currents lying below the mass bounds  in (\ref{mass_bounds}).
However, the compatibility of the mass bounds with the fitted values of Table \ref{best_tab}
are useful as a consistency check on the fitting
procedure. The bounds also provide a constraint on single-resonance scenarios which is valid for
a wide class of resonance shapes.

\subsection{$\pi\pi$ Scattering and the $\sigma$  Resonance Shape}
\label{shape_sec}
The underlying idea in sum-rule methodology is to compare the imaginary parts of the correlation 
function obtained from QCD  with the imaginary parts of the propagator of the resonance under investigation, 
{\it i.e.}
\begin{equation}
\frac{1}{m_r^2-s-im_r G'}\quad ,
\label{prop}
\end{equation}
where $G'$ is a regularization parameter.  A constant (energy independent) $G'$ 
corresponds to a Breit-Wigner description of the resonance.  In general, however, 
$G'$ is energy dependent [$G'=G'(s)$] and can  be determined from the dynamics by which the resonance is probed.
For example, the $\sigma$ resonance shape as probed by $\pi\pi$ scattering has been evaluated in
ref. \cite{Sannino}:
\begin{eqnarray}
& &\frac{m_\sigma G(s)}{m_\sigma^2-s-im_\sigma G'}
\quad ,
\label{amir1}\\
& &m_\sigma G(s)=\frac{3 g^2_{\sigma\pi\pi}\sqrt{s-4m_\pi^2}}{64\pi\sqrt{s}}\left(s-2m_\pi^2\right)^2
\quad ,
\label{amir2}
\end{eqnarray}
with $\Gamma_\sigma=G(m_\sigma^2)$.  
Following the general prescriptions for construction of unitary scattering amplitudes given in 
\cite{new_ahf}, 
unitarity of this scattering amplitude at {\em any} point $s$ 
implies $G'=G(s)$.  This determines the $\sigma$ contribution to the hadronic correlation function in the scalar channel:
\begin{equation}
\Pi(s)=g\frac{1}{m_\sigma^2-s-im_\sigma G'}
\quad ,
\label{sigma_corr}
\end{equation}
where $g$ adjusts the dimensionality of the hadronic and field-theoretic correlation function.
Therefore, substituting $G'=G(s)$, as given by (\ref{amir2}) we find that 
\begin{equation}
Im\Pi^{res}(s)=g
\frac{m_\sigma\Gamma_\sigma F_\sigma\sqrt{s}\sqrt{s-4m_\pi^2}\left(s-2m_\pi^2\right)^2}{s\left(s-m_\sigma^2\right)^2+m_\sigma^2\Gamma_\sigma^2F_\sigma^2\left(s-4m_\pi^2\right)\left(s-2m_\pi^2\right)^4}
\quad ,
\label{sig_res}
\end{equation}
where
\begin{equation}
F_\sigma=\frac{m_\sigma}{\left(m_\sigma^2-2m_\pi^2\right)^2\sqrt{m_\sigma^2-4m_\pi^2}}\quad .
\label{F_sigma}
\end{equation}
Obviously this resonance shape vanishes in the limits $s\rightarrow\infty$ and $\Gamma_\sigma\rightarrow 0$,
as required.  When $m_\pi\rightarrow 0$ we see that this resonance shape may be expressed as
\begin{equation}
\frac{1}{\pi}Im\Pi^{res}(s)=\frac{G_r}{M^2}\frac{s^2}{\left(s-M^2\right)^2+\frac{\Gamma^2}{M^6}s^4}
\quad .
\label{new_res}
\end{equation}
Since this resonance shape is asymmetric, it could in principle violate the bounds (\ref{mass_bounds}).
The resonance shape (\ref{new_res}) leads to the following  resonance contribution 
to the sum-rule $R_0$:
\begin{eqnarray}
& &\int\limits_0^{s_0} e^{-s\tau} \frac{1}{\pi}Im\Pi^{res}(s) \, ds\equiv G_r
e^{-M^2\tau}W_0(M,\Gamma,\tau,s_0)\quad ,
\label{new_res_sr}\\
& & W_0(M,\Gamma,\tau,s_0)=\int\limits_{-1}^{\frac{s_0-M^2}{M^2}} 
\frac{e^{-\xi M^2\tau}(\xi+1)^2}{\xi^2+\frac{\Gamma^2}{M^2}(\xi+1)^4}
\, d\xi\quad ,
\label{new_w}\\
& &R_0(\tau,s_0)=G_re^{-M^2\tau}W_0(M,\Gamma,\tau,s_0)\quad .
\label{new_sr}
\end{eqnarray}

Equations (\ref{new_res_sr}--\ref{new_sr}) provide a means for comparing the field-theoretical content 
(\ref{laplace_theor}) of $R_0$ in the isoscalar channel with the phenomenological $\sigma$-resonance
shape obtained in ref. \cite{Sannino}.
Defining $A=G_r/\hat m^2$ and performing a $\chi^2$ minimization 
[see (\ref{chi2})]
with this new resonance model leads to the best-fit parameters shown in
Table \ref{new_fit_tab}.  Since the mass prediction in Table
\ref{new_fit_tab}
is very close to the mass bound (\ref{mass_bounds}), the new resonance shape
is able to accommodate asymmetric width effects without any strong violation of the
lower bound.

\begin{table}[hbt]
\begin{tabular}{||c|c|c|c||}\hline\hline
$M~(GeV)$ & $s_0~(GeV^2)$ & $A~(GeV^4)$ & $\Gamma ~(GeV)$
\\ \hline\hline
  $0.86$ & $4.1$ & $0.082$ & $0.34$ 
\\\hline\hline
\end{tabular}
\caption{  
 Values
 for the resonance parameters in (\protect\ref{new_res_sr})
and continuum threshold for the $I=0$ channel
which lead to a minimum $\chi^2$.
}
\label{new_fit_tab}
\end{table}

Results for the Monte-Carlo simulation of uncertainties (see Section \ref{one_res_sec}) are
shown in Table \ref{new_mc_tab}.  These results again indicate that a very
light ($400$--$700\,{\rm MeV}$) $\sigma$ cannot be readily identified with the QCD prediction for the
$I=0$ $n\bar n$ state.

\begin{table}[hbt]
\begin{tabular}{||c|c|c|c||}\hline\hline
$M~(GeV)$ & $s_0~(GeV^2)$ & $A~(GeV^4)$ & $\Gamma ~(GeV)$
\\ \hline\hline
  $0.86\pm 0.07 $ & $4.01\pm 0.43$ & $0.091 \pm 0.033$ & $0.33\pm 0.09$ 
\\\hline\hline
\end{tabular}
\caption{  
Results of the Monte-Carlo simulation of
90\% confidence level uncertainties  for the resonance parameters in (\protect\ref{new_res_sr})
and continuum threshold for the $I=0$ channel.
}
\label{new_mc_tab}
\end{table}

\section{Conclusions}
\label{conc_sec}
The self-consistency of the QCD Laplace sum-rules which probe the non-strange $n\bar n$ component of the $I=\{0,1\}$ 
scalar 
mesons has been studied using H\"older inequality techniques.  This analysis
confirms the methodological consistency of including an instanton continuum contribution \cite{instanton_continuum} 
which increases the instanton contributions, and hence could enhance the 
isospin splitting between the $I=\{0,1\}$ channels beyond that  observed in ref. \cite{first_paper} to accommodate a light $\sigma$.  

Motivated by this possible enhancement, we have conducted 
an extensive phenomenological analysis of the Laplace and finite-energy sum-rules for the
non-strange $n\bar n$ component of the scalar resonances. 
Theoretical uncertainties resulting from resonance-shape effects, multiple resonance contributions,
and higher-loop perturbative contributions have also been considered.  
The results indicate that  
neither  $\sigma$ nor  $a_0(980)$  dominate the sum-rules. Instead they appear to be weakly coupled 
to the $n\bar n$ current (\ref{current})
compared with a  $M\approx 1\,{\rm GeV}$ isoscalar and a $M\approx 1.5\,{\rm GeV}$ isovector which  
are consistently seen to dominate the sum-rule analysis.  
The decoupling of a light $\sigma$ and the $a_0(980)$ from the  $n\bar n$
scalar sum-rules suggests  a non-$q\bar q$  interpretation for these resonances.
 Furthermore, we estimate  the $n\bar n$ content of the $a_0(980)$ to be at most $\sim 20\%$  
and that of a $500\,{\rm MeV}$  $\sigma$
to be at most $\sim 30\%$. 
These results certainly allow room for a meson-meson or four-quark interpretation of $a_0(980)$
and $\sigma(\sim 500)$; the $n\bar n$ components of these states are clearly 
secondary.

 Thus the $n\bar n$ scalar-channel sum rules argue strongly against interpreting either  
$a_0(980)$ or  an order-$500\,{\rm MeV}$ $\sigma$ 
as relatively pure $n\bar n$ states.  
The results for these states contrast strongly against 
 the clear signature  for an $n\bar n$-interpretation 
of the $\rho$ in the context of vector-channel sum rules, or of the pion (and even the
first pion excitation state) in the context of pseudoscalar sum rules \cite{Hubschmid,SVZ,Steele}.

Consequently, the analysis presented here does not support interpretations that include the $a_0(980)$
and a $\sim 500\,{\rm MeV}$ $\sigma$ in a primitive $q\bar q$ scalar nonet.
 It is also important to note that our analysis clearly distinguishes between the $f_0(980)$ and $a_0(980)$, both of which might
be expected to contain a substantial $K\bar K$ component on the basis of their proximity to the $K\bar K$ threshold.  This distinction
between the $I=0$ and $I=1$ states is also a feature of models of the $K\bar K$ interaction \cite{Speth}.

Sum-rule analysis is in excellent agreement with the identification of the $a_0(1450)$ as the lightest $I=1$
quark ($q\bar q$) scalar resonance with  significant $n\bar n$ content, 
supporting the conclusions of refs. \cite{nonet,new_ahf_2}.  The situation is not as clear for the isoscalar channel, since 
uncertainty in the predicted width and
mass (both from theoretical uncertainties and resonance model dependence) could accommodate a dominant state 
anywhere between
$800\,{\rm MeV}$ and $1100\,{\rm MeV}$, including  the $f_0(980)$,  as the lightest 
quark ($q\bar q$) scalar meson with a significant $n\bar n$ content.

\smallskip
\noindent
{\bf Acknowledgements:}  TGS, VE and FS are grateful for research support
from the Natural Sciences and Engineering Research Council of Canada (NSERC).
Work of AHF has been supported in part by the
US DOE under contract DE-FG-02-85ER-40231.
 AHF gratefully acknowledges
discussions with J. Schechter and D. Black.
TGS thanks M. Pennington and K. Maltman for stimulating discussions.

\end{document}